\documentclass{aa}
\usepackage{epsfig,psfig}
\usepackage{txfonts}

\usepackage{epsfig}

\newcommand{\Msun}{\mbox{$M_{\odot}$}}

\newcommand{\logl}{\mbox{$\log (L/L_{\odot}$)}}  

\newcommand{\zsun}{\mbox{$Z_{\odot}$}}

\newcommand{\kms}{km s$^{-1}$}
\newcommand{\degree}{$^{\rm o}$}

\newcommand{\bb}{\bibitem[]{bla}}

\begin{document}


\title{Wolf-Rayet spin at low metallicity and its implication\\
for Black Hole formation channels}

\author{Jorick S. Vink\inst{1}, Tim J. Harries\inst{2}}
\offprints{Jorick S. Vink, jsv@arm.ac.uk}

\institute{Armagh Observatory and Planetarium, College Hill, Armagh BT61 9DG, Northern Ireland, UK
          \and
          Department of Physics and Astronomy, University of Exeter, Exeter EX4 4QL, UK}

\titlerunning{Spin and wind asphericity of low-$Z$ Wolf-Rayet stars}
\authorrunning{Jorick S. Vink}

\abstract {The spin of Wolf-Rayet (WR) stars at low metallicity ($Z$) is most relevant for our understanding 
of gravitational wave sources such as GW\,150914, as well as the incidence of long-duration gamma-ray bursts (GRBs). Two scenarios
have been suggested for both phenomena: one of them involves rapid rotation and quasi-chemical homogeneous evolution (CHE), the other 
invokes classical evolution through mass loss in single and binary systems. }
{The stellar spin of Wolf-Rayet stars might enable us to test these two scenarios.
In order to obtain empirical constraints on black hole progenitor spin 
we infer wind asymmetries in all 12 known WR stars in the Small Magellanic Cloud (SMC) at $Z=1/5\zsun$, as well as 
within a significantly enlarged sample of single and binary WR stars in the Large Magellanic Cloud (LMC at $Z=1/2\zsun$), tripling the 
sample of Vink (2007). This brings the total LMC sample to 39, making it appropriate for comparison to the Galactic sample.}
{We measure WR wind asymmetries with VLT-FORS linear spectropolarimetry, a tool uniquely poised to 
perform such tasks in extra-galactic environments.}
{We report the detection of new line effects in the LMC WN star BAT99-43 and the WC star BAT99-70, as well as the 
famous WR/LBV HD\,5980 in the SMC, which might be evolving chemically homogeneously. 
With the previous reported line effects in the late-type WNL (Ofpe/WN9) 
objects BAT99-22 and BAT99-33, this brings the total LMC 
WR sample to 4, i.e. a frequency of $\sim$10\%.
Perhaps surprisingly, the incidence of line effects amongst low $Z$ WR stars is not 
found to be any higher than amongst the Galactic WR sample, challenging the rotationally-induced CHE model.}
{As WR mass loss is likely $Z$-dependent, our Magellanic Cloud 
line-effect WR stars 
may maintain their surface rotation and fulfill the basic conditions for producing 
long GRBs, both via the  
classical post-red supergiant (RSG) or luminous blue variable (LBV) channel, as well as resulting from CHE due to physics specific to very massive stars (VMS).}


\maketitle

\section{Introduction}
\label{s_intro}

Long-duration gamma-ray bursts (GRBs) are amongst the brightest and most 
distant objects in the Universe (Tanvir et al. 2009; Cucchiara et a. 2011), 
implying that massive stars could live and die when the cosmos was just a few hundred million years old.
The favored progenitors are Wolf-Rayet (WR) stars at low metallicity (Woosley \& Bloom 2006).
In the collapsar scenario (MacFadyen \& Woosley 1999) a rapidly rotating stellar 
core collapses into a black hole, thereby producing two narrow jets that are identifiable as a GRB. 
As of today the progenitors of GRBs are poorly constrained empirically, and one of the aims of our 
WR studies in low metallicity ($Z$) galaxies is to alleviate this shortcoming. 
The key physical sequence of events with respect to the GRB puzzle concerns 
the interplay between mass loss and rotation as a function of metallicity ($Z$), with the main idea 
that lower $Z$ WR stars have weaker stellar winds than their galactic counterparts (Vink \& de Koter 2005). 
This should lead to less spindown of the progenitor WR star, enabling GRB conditions at low $Z$, e.g. 
via rotation-induced chemically-homogeneous evolution (Yoon \& Langer 2005; Woosley \& Heger 2006).

A second major motivation for our study concerns the discovery of huge black hole masses in the merging
event associated with GW150914, involving masses of order 30-40$\Msun$ (Abbott et al. 2016). This has led 
to a  
quest for constraints on the spin rates of black hole progenitors, i.e. WR stars. 
Given the huge masses inferred in the GW\,150914 event, it is most likely that these extremely 
massive binary black holes were situated in a low-$Z$ environment, as this is where the maximum black 
hole masses are predicted to be larger (Belczynski et al. 2010). 
It is therefore relevant to constrain the rotation rates of both single and binary 
WR stars in the low metallicity LMC ($Z=1/2\zsun$) and SMC ($Z=1/5\zsun$) galaxies. 

We note that the evolution of massive stars into WR stars and subsequently into black holes
is still very uncertain, even for single star evolution. It is as yet not known whether 
internal magnetic fields couple the core to the envelope (e.g. Brott et al. 2011) or if massive stars
rotate differentially (e.g. Georgy et al. 2013). For the former case, one may expect solid-body
rotation, where the surface rotation rate inferred from observations may provide direct
information on the spin properties of black hole progenitors. In the latter case, such inferences
would be less direct, although relevant constraints on the core rotation of Wolf-Rayet stars may 
still be obtained.

The special property of WR stars in comparison to canonical stars is that all their lines are in emission.
This means that the traditional method involving the width of absorption lines to measure the 
rotation rate (actually v$\sin i$) does not work for WR stars, but thanks to the line emission\footnote{Note that Shenar et al. (2014) have calculated
the potential influence of rotation for WR emission line shapes.} an alternative 
tool is available to measure wind asymmetry resulting from rapid rotation.
In its simplest form, the tool is based on the expectation that line photons arise 
over a larger volume than continuum photons, such that line photons undergo 
fewer scatterings and the emission-line flux is less polarized than the continuum
when the wind geometry is aspherical when projected on the sky. 
This results in a smooth polarization variation across the line profile: the ``line effect''. 

The high incidence of line effects amongst classical Be stars (Poeckert \& Marlborough 1976) 
indicated that the envelopes of classical Be stars are not spherically symmetric.  
In a similar vein Harries et al. (1998) performed spectropolarimetry on a sample  
of 29 Galactic WR stars, finding that in the vast majority (of 80\%) of WRs 
line effects were absent, indicating that the winds are spherical and rapid rotation is rare. 
Vink et al. (2011) and Gr\"afener et al. (2012) noted that the exceptions amongst the Galactic 
WRs were not 
randomly distributed but that there exists a highly significant correlation between line-effect 
WR stars and WRs with ejecta nebula, arguing that it is only the ``young'' WR stars -- that have only
recently transitioned from a previous red supergiant (RSG) or luminous blue variable (LBV)
phase -- that rotate rapidly, presumably before strong Galactic WR winds during the 
rest of the WR phase ensure that the bulk of Galactic WR stars is spherical, rotating slowly.
This suggests a post RSG or LBV
``classical'' scenario for the production of GRBs rather than quasi-chemical homogeneous evolution (CHE), as discussed 
in Vink et al. (2011) and Gr\"afener et al. (2012).

As WR winds in low metallicity environments are thought to be weaker (Vink \& de Koter 2005; Gr\"afener \& Hamann 2008; Hainich et al. 2015) one may possibly expect a higher incidence of line 
effect WR stars in the LMC and SMC. Vink (2007) therefore performed linear spectropolarimetry on 
a sample of the brightest 12 WRs in the LMC, discovering only 2 line effects in 
BAT99-22 and BAT99-33, which is similar to the incidence rate in the Galaxy. 
Here we extend our study to the even lower metallicity of the SMC, and we 
also extend the LMC sample to a sample size comparable to that of the Galaxy.
The sample of Vink (2007) was necessarily biased towards the brightest 
objects (with $V$ $<$ 12, mostly containing very late-type nitrogen-rich WR stars and/or binaries),
and for an unbiased assessment one requires a larger sample.
By encompassing the magnitude range $V$ $<$ 14, we triple the LMC sample (from 13 to 39). 

Although both WN (nitrogen-rich) and WC (carbon-rich) stars might eventually 
become GRBs, WC stars are more likely to be the {\it direct} progenitors 
(e.g. Levan et al. 2016). Nevertheless, the physics of enhanced rotation is equally 
relevant for both sub-groups. Moreover, it is hard to tell whether WR stars in all mass ranges 
(and for all rotation rates and metallicities) always evolve from WN into WC stars (see e.g. Sander et al. 2012). 
For these reasons, we wish to constrain wind asymmetries and rotation rates 
amongst both sub-groups of WR stars.
Furthermore, we wish to correlate linear polarization and rapid rotation with binarity.
The larger data-set enables us to correlate linear polarization 
with binarity, as radial velocity (RV) surveys for binary WRs in the LMC 
are available for both WC and WN stars respectively (Bartzakos et al. 2001; Foellmi et al. 2003), 
allowing us to address the question whether binarity is a prerequisite for observed 
linear polarization in WR stars. 

The paper is organized as follows. 
In Sect.~\ref{s_obs} we briefly describe the observations, data reduction, and analysis
of the linear polarization data. This is followed by a description of the resulting 
line Stokes $I$ and linear polarization profiles for the SMC and LMC WR stars in Sect.~\ref{s_res}. 
In Sect.~\ref{s_disc}, we discuss the constraints
these observations provide on WR spin rates and evolution, and GRB production models.

\section{Observations, data reduction, and methodology}
\label{s_obs}

The SMC linear spectropolarimetry data were obtained during the 
nights of 2008 September 8,9, and 10 during ESO Period 81 
using the FORS spectrograph in PMOS mode. 
Apart from the grism, the data were obtained in a similar manner as for 
the brighter ($V < 12$) LMC sample of Vink (2007) during P78. 
Here we use the 300V grism and the GG375 order filter with a slit width 
of 2$\arcsec$, yielding a spectral resolution of approximately 
20\AA. This spectral resolution is about three times lower than that 
of Vink (2007) for the LMC and Harries et al. (1998) for the Milky Way measurements.

The new LMC data were obtained in an almost identical way as our SMC data, 
during the nights of 2010 September 22-25 (ESO Period 85).
Our LMC targets were selected from the fourth catalogue of Population I LMC WR stars 
by Breysacher, Azzopardi \& Tester (1999; hereafter BAT) on the basis of their 
relative brightness ($V \la 14$). Note that line-effect stars in the new LMC sample are 
seen down to V magnitudes of 14, suggesting that brightness is not a pre-requisite for showing polarization.
In order to obtain a sample as unbiased as possible -- save for brightness -- the objects were not selected
on the basis of any known circumstellar (CS) geometries, spectral peculiarities, or binarity.
The list of objects is given in Table~\ref{t_cont}, alongside their spectral types.
Because the 26 new LMC objects are fainter than the 13 brighter LMC objects observed in cycle 78 by Vink (2007), we settle for 
lower spectral resolution, which should not pose any problems as WR stars 
have fast winds (of thousands km/s).

To analyze the linearly polarized component in the spectra, FORS was equipped 
with the appropriate polarization optics.
Polarization and un-polarized standard stars were also observed.
Data reduction consisted of the preparation of the images via the subtraction 
of a median bias frame and the construction of a median flat-field for each night. 
The flat-field was then normalized via a division with a smoothed copy, and this normalized 
flat-field was subsequently divided into the target images. Wavelength calibrations were 
constructed using low-order polynomial fits to the dispersion curve defined by visually identified
arc lines. The spectropolarimetric reduction then closely followed the method detailed by 
Harries \& Howarth (1996), although we note no cross-talk correction was required since the 
ordinary ($o$) and extraordinary ($e$) spectra are well separated on the FORS images. 
The $o$ and $e$ spectra were first extracted and sky subtracted, wavelength calibrated and 
rebinned onto a uniform wavelength array. The 8 spectra ($o$ and $e$ spectra from the
0$^\circ$, 45$^\circ$, 22.5$^\circ$, and 67.5$^\circ$, images) were finally combined 
into Stokes $I$, $Q$, and $U$ intensity spectra via the ratio method (e.g. Tinbergen 1996), 
formally propagating the shot-noise errors throughout in order to provide variances on $Q$ and $U$.

Observations of an unpolarized standard (HD\,10038) were taken during the LMC run, and these indicated that the instrumental polarization was less than
0.1\%. We also obtained spectra of polarized standard stars (NGC\,2024, Hiltner\,652, and BD-14\,1922) and the polarization magnitudes of these objects
showed good agreement with the literature values. However, the position angles (PAs) showed a small, consistent offset of $2^\circ$ (relative to the
literature values). More importantly, all spectra showed a slow PA fluctuation with wavelength. We found this PA rotation to be the same for all our
polarized standards and it is attributable to a slight chromatism of the half-wave plate. We fitted this PA rotation with a 5th-order polynomial, under the
pragmatic assumption that the PAs of the polarized standards are constant over the wavelength range of the data. We then used this polynomial fit to
correct the polarization spectra of our targets. 

The percentage linear polarization $P$ and its position angle $\theta$ are determined in the following way:

\begin{equation}
P~=~\sqrt{(Q^2 + U^2)}
\end{equation}
\begin{equation}
\theta~=~\frac{1}{2}~\arctan(\frac{U}{Q})
\end{equation}
The achieved accuracy of the polarization data is in principle determined by 
photon-statistics only, however due to 
systematic effects the absolute accuracy might be lower. We do not correct 
for instrumental or interstellar polarization (ISP) as these are equal for line and continuum 
wavelengths. 
Depolarization line effects can be measured across the 
He {\sc ii} lines at 4686\AA, 6560\AA, as well as several other emission lines, such as the 
the C{\sc iv} line at 5805\AA\ for the WC stars. 
As WR stars have fast winds (of order thousands \kms), we easily resolve these lines for most targets\footnote{Note that 
some of the targets have slower winds, and these line effects are thus unresolved. However, given that WR line effects are 
known to be consistent with simple depolarizations (rather than more subtle line polarization effects, e.g. Vink et al. (2002), we do not 
expect this to affect our results.}. 

We note that a non-detection would imply the wind is
circularly symmetric on the sky (to within the
detection limit). The detection limit is inversely dependent on the
signal-to-noise ratio (SNR) of the spectrum, and the contrast of the emission line as the
line-emission is diluting/depolarizing the flux from the continuum. Following Davies et al. (2005), 
the detection limit for the maximum intrinsic polarization $\Delta P_{\rm limit}$ can be represented 
by:

\begin{equation}
\Delta P_{\rm limit} (\%) = \frac{100}{SNR} \times \frac{l/c}{l/c-1}
\label{eq:pint}
\end{equation} 

\noindent where $l/c$ refers to the line-to-continuum contrast. 

\begin{table*}
\caption{Polarization data of WR stars observed during the various VLT runs. The brighter 13 LMC stars from Vink (2007) are at the bottom.} 
\label{t_cont}
\begin{tabular}{llllcrcrcc}
\hline
Name & Alt. names & Spec.Tp & $V$ mag  & $P_{\rm cont}^{\rm B}$ (\%) & $\theta_{\rm cont}^{\rm B}$ (\degree) & $P^{\rm lit}$ (\%) & $\theta^{\rm lit}$ & line effect & binarity\\
\hline
\\
{\it SMC}:\\
\\
SMC-WR1 & AV 2a         & WN3     & 15.14 & 0.922 $\pm$ 0.03 &  41.0 $\pm$ 0.8 &      &     &   & no \\
SMC-WR2 & AV 39a        & WN4.5   & 14.23 & 0.196 $\pm$ 0.02 & 154.0 $\pm$ 2.9 &      &     &   & no \\
SMC-WR3 & AV 60a        & WN3     & 14.48 & 0.701 $\pm$ 0.02 & 131.0 $\pm$ 0.9 &      &     &   & yes\\
SMC-WR4 & AV 81, Sk 41  & WN6p    & 13.35 & 0.416 $\pm$ 0.02 & 126.0 $\pm$ 1.3 &      &     &   & no\\
SMC-WR5 & HD 5980       & WN5     & 11.08 & 0.117 $\pm$ 0.02 &  27.6 $\pm$ 4.7 &      &     & x & yes\\
SMC-WR6 & AV 332, Sk 108& WN3     & 12.30 & 0.416 $\pm$ 0.01 & 122.2 $\pm$ 0.8 &      &     &   & yes\\
SMC-WR7 & AV 336a       & WN2     & 12.93 & 0.263 $\pm$ 0.02 & 119.7 $\pm$ 1.8 &      &     &   & yes\\
SMC-WR8 & Sk 188        & WO4     & 12.81 & 0.221 $\pm$ 0.02 & 103.2 $\pm$ 2.2 &      &     &   & yes\\
SMC-WR9 &               & WN3     & 15.23 & 0.437 $\pm$ 0.03 & 132.0 $\pm$ 1.6 &      &     &   & ?\\
SMC-WR10&               & WN3     & 15.76 & 0.458 $\pm$ 0.03 & 167.5 $\pm$ 2.0 &      &     &   & no\\
SMC-WR11&               & WN3     & 14.97 & 0.408 $\pm$ 0.03 & 139.2 $\pm$ 2.4 &      &     &   & no\\
SMC-WR12& SMC-054730    & WN3     & 15.46 & 0.922 $\pm$ 0.03 &  41.0 $\pm$ 0.8 &      &     &   & no \\
\hline
\\
{\it LMC BAT99}:\\
\\
7  & & WN4b       & 14.10 & 0.888 $\pm$ 0.022 &   34.3 $\pm$ 0.7&                 &          &  & no\\
8  & & WC4        & 14.08 & 0.789 $\pm$ 0.029 &   34.5 $\pm$ 1.0&                 &          &  & no?\\
9  & & WC4        & 14.05 & 1.164 $\pm$ 0.045 &   31.9 $\pm$ 1.1&                 &          &  & no?\\
10 & R 64 & WC4$+$O9.5 & 13.61 & 0.417 $\pm$ 0.011 &   66.1 $\pm$ 0.7 & 0.66 $\pm$ 0.09 & 76 $\pm$ 4 & & no\\
11 & & WC4        & 13.01 & 0.413 $\pm$ 0.043 &   23.6 $\pm$ 3.0 &                 &          & & no  \\
14 & & WN4o$+$OB  & 13.7 & 0.651 $\pm$ 0.026 &   30.6 $\pm$ 1.1 &                 &           & & no\\
19 & & WN4b$+$O5  & 13.76 & 0.052 $\pm$ 0.020 &    3.3 $\pm$11.0 &                 & & & yes            \\
20 & & WC4$+$O    & 14.01 & 2.322 $\pm$ 0.031 &  178.3 $\pm$ 0.4 &                 &   & & no         \\
21 & & WN4o$+$OB  & 13.11 & 0.632 $\pm$ 0.015 &   32.3 $\pm$ 0.7 &                 &   & & no        \\
34 & & WC4$+$OB   & 12.72 & 0.404 $\pm$ 0.020 &   31.5 $\pm$ 1.4 &                 &       &   & yes?     \\
43 & & WN4o$+$OB  & 14.18 & 0.611 $\pm$ 0.019 &   10.1 $\pm$ 0.9 &                 &       & x & yes   \\
47 & & WN3b       & 14.11 & 0.516 $\pm$ 0.033 &   28.0 $\pm$ 1.9 &                 &  & & no\\
52 & & WC4        & 13.37 & 0.247 $\pm$ 0.030 &   30.2 $\pm$ 3.5 &                 &  & & no\\
53 & & WC4$+$OB   & 13.17 & 0.565 $\pm$ 0.019 &   34.6 $\pm$ 1.0 &                 &   & & yes?\\
59 & & WN4b$+$O8  & 13.33 & 0.188 $\pm$ 0.014 &   34.3 $\pm$ 2.2 & 0.17 $\pm$ 0.27 & 111$\pm$ 29 & & ?\\
61 & & WC4        & 13.04 & 0.301 $\pm$ 0.025 &   35.0 $\pm$ 2.3 & 0.27 $\pm$ 0.10 & 31 $\pm$ 11 & & no\\
64 & & WN4o$+$O9  & 14.39 & 0.538 $\pm$ 0.018 &   49.7 $\pm$ 1.0 &                 & & & yes  \\
67 & & WN5ha      & 13.89 & 1.256 $\pm$ 0.024 &   36.2 $\pm$ 0.6 &                 &  & & no\\
70 & Sk -69 207 & WC4$+$OB   & 13.79 & 2.205 $\pm$ 0.045 &    0.0 $\pm$ 0.6 & 0.95 $\pm$ 0.13 & 19 $\pm$ 4 & x & no\\
84 & & WC4$+$OB   & 13.00 & 0.399 $\pm$ 0.020 &  164.1 $\pm$ 1.4 & 0.91 $\pm$ 0.20 & 171$\pm$ 7 &   & no\\
92 & R 130 & WN6$+$B1   & 11.51 & 0.943 $\pm$ 0.011 &   87.9 $\pm$ 0.3 &                 &            \\
117& R 146 & WN5ha      & 12.99 & 1.025 $\pm$ 0.021 &   38.5 $\pm$ 0.6 & 0.70 $\pm$ 0.26 & 48 $\pm$ 10 & & no\\
122& R 147 & WN5(h)     & 13.06 & 1.487 $\pm$ 0.020 &   32.3 $\pm$ 0.4 & 1.55 $\pm$ 0.18 & 31 $\pm$ 3  & & no\\
125& & WC4$+$OB   & 13.18 & 1.164 $\pm$ 0.016 &  166.7 $\pm$ 0.4 &                 &           & & yes?\\
126& & WN4b$+$O8  & 13.35 & 0.932 $\pm$ 0.015 &  169.2 $\pm$ 0.5 &                 &     & & ?        \\
127& & WC5$+$O6   & 13.31 & 0.818 $\pm$ 0.014 &  163.6 $\pm$ 0.5 &                 &  & & yes?           \\
\hline
22 & R 84    &  WN9h    & 12.09       &  0.235 $\pm$ 0.007      &  147.8 $\pm$ 0.9  & 0.30 $\pm$ 0.15 & 74 $\pm$13       & x & \\       
27 &      &  WN5?b$+$B1Ia   & 11.31        &  0.254 $\pm$ 0.006      &   36.3 $\pm$ 0.7  & 0.22 $\pm$ 0.13  & 73 $\pm$ 15 & & no\\
28 & R 90 &  WC6$+$O5-6V-III     & 12.23        &  1.038 $\pm$ 0.007      &   47.5 $\pm$ 0.2 & 0.77 $\pm$ 0.17 & 53 $\pm$ 6 & & yes  \\
33 & R 99   &  Ofpe/WN9       & 11.54        &  1.315 $\pm$ 0.007      &  106.2 $\pm$ 0.2 & &       & x &       \\
38 &  &  WC4$+$O?           & 11.50        &  0.556 $\pm$ 0.007      &   21.1 $\pm$ 0.3  & 0.48 $\pm$ 0.15 & 16 $\pm$ 9 & & yes  \\
39 &   &  WC4$+$O6V-III    & 12.51        &  0.445 $\pm$ 0.013      &   23.0 $\pm$ 0.9   & & & & yes            \\
42 & R 103  &  WN5?b$+$(B3I)    &  9.91        &  0.601 $\pm$ 0.007      &   24.5 $\pm$ 0.4              & & no  \\
  55 &  &  WN11h          & 11.99        &  0.226 $\pm$ 0.009      &   30.7 $\pm$ 1.1           \\
  85 &  &  WC4$+$OB        & 11.75        &  1.716 $\pm$ 0.007      &  104.1 $\pm$ 0.1   & 1.59 $\pm$ 0.09 & 99 $\pm$ 2 & &no    \\
  92 & R 130  &  WN6$+$B1Ia      & 11.51        &  1.043 $\pm$ 0.006      &   81.6 $\pm$ 0.2             \\
 107 & R 139 &  WNL/Of         & 12.12        &  1.647 $\pm$ 0.006      &   70.1 $\pm$ 0.1             \\
 118 & R 144 &  WN6h           & 11.15        &  0.166 $\pm$ 0.006      &   20.6 $\pm$ 1.1  & 0.05 $\pm$ 0.14 & 48 $\pm$ 10             \\
 119 & R 145 &  WN6(h)         & 12.16        &  2.231 $\pm$ 0.006      &   81.2 $\pm$ 0.1  & 1.82 $\pm$ 0.23 & 79 $\pm$ 4             \\
\hline

\end{tabular}
\\
\noindent
Names, spectral types, and $V$ magnitudes (columns 1-4) are from Massey \& Duffy (2001) and Massey et al. (2003) for the SMC, 
and from Breysacher et al. (1999, BAT99) for the LMC.
The continuum polarization percentage and errors (over 4200-4600\AA) are in column 5. 
Systematic errors in the polarization are estimated to be of order 0.1\%. 
PAs ($\theta$) are listed in column 6. 
The literature values of the $B$-band PA and continuum polarization are from the
catalogue of Mathewson \& Ford (1970) (columns 7 and 8). 
Binary information is given in column 10, for the SMC from Shenar et al. (2016), and 
for the LMC from Bartzakos et al. (2001) for WCs and Foellmi et al. (2003) for WNs.
\end{table*}

Our final spectra had between 10$^5$ and 10$^6$ counts per spectral bin in the continuum,
giving a polarization error of between $\sim 0.3$\% and $\sim 0.1$\% and the spectral resolution, measured from the arc spectra, was $\sim 20$\AA, 
corresponding to $\sim$1000 km/s. 
In Figs. 1 - 4 we plot the calibrated polarization spectra of our targets binned to a constant error of 0.1\% in polarization.

\section{Results}
\label{s_res}

\begin{figure*}
\mbox{
\epsfxsize=0.25\textwidth\epsfbox{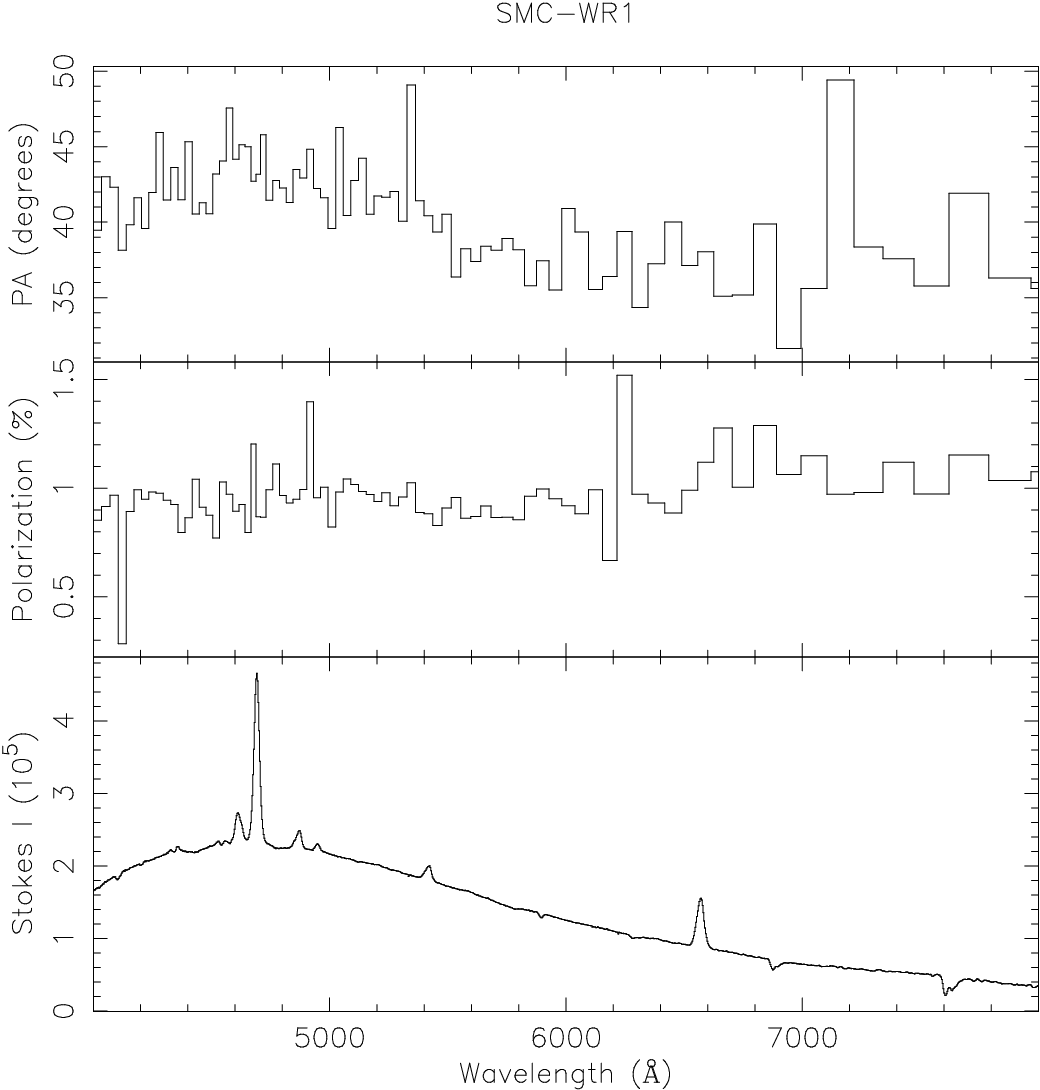}
\epsfxsize=0.25\textwidth\epsfbox{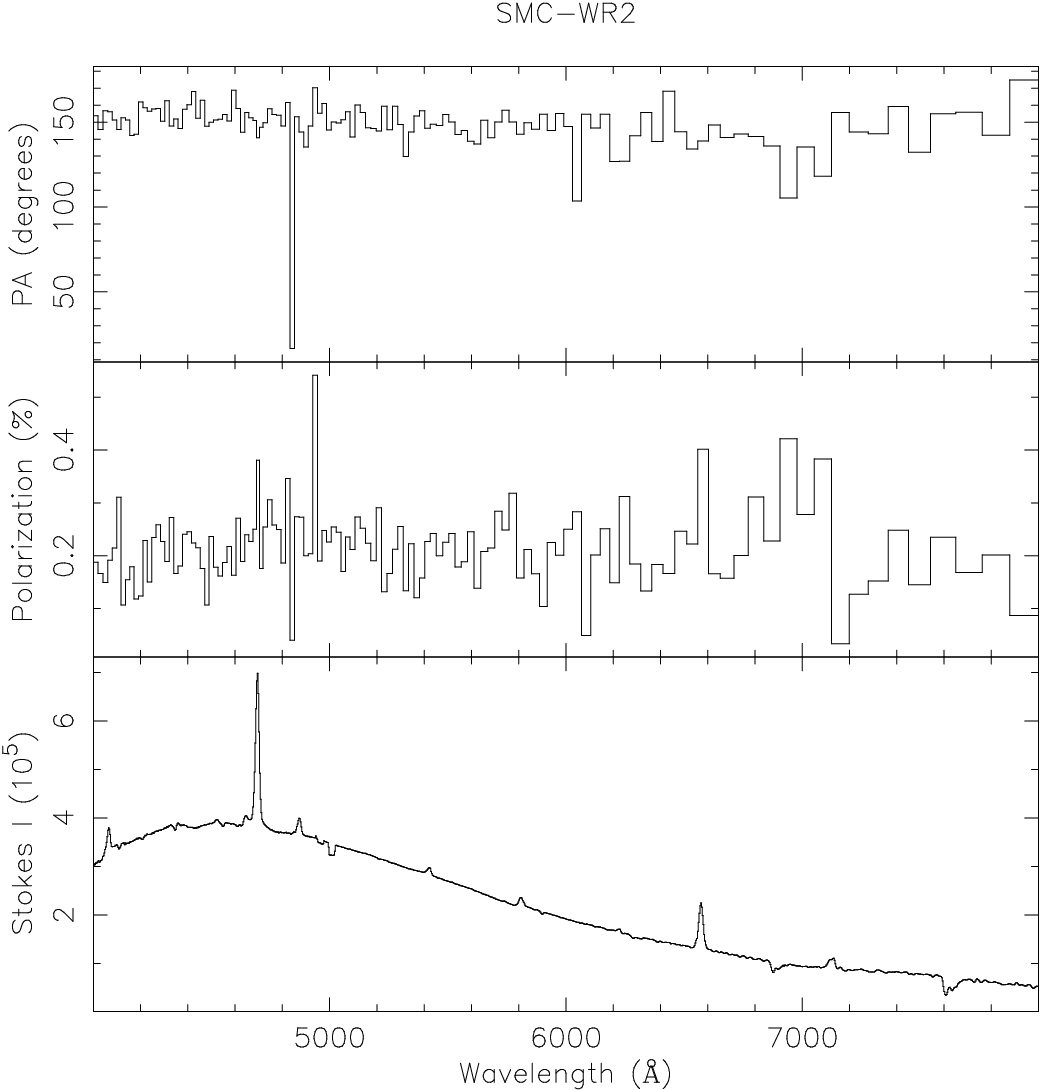}
\epsfxsize=0.25\textwidth\epsfbox{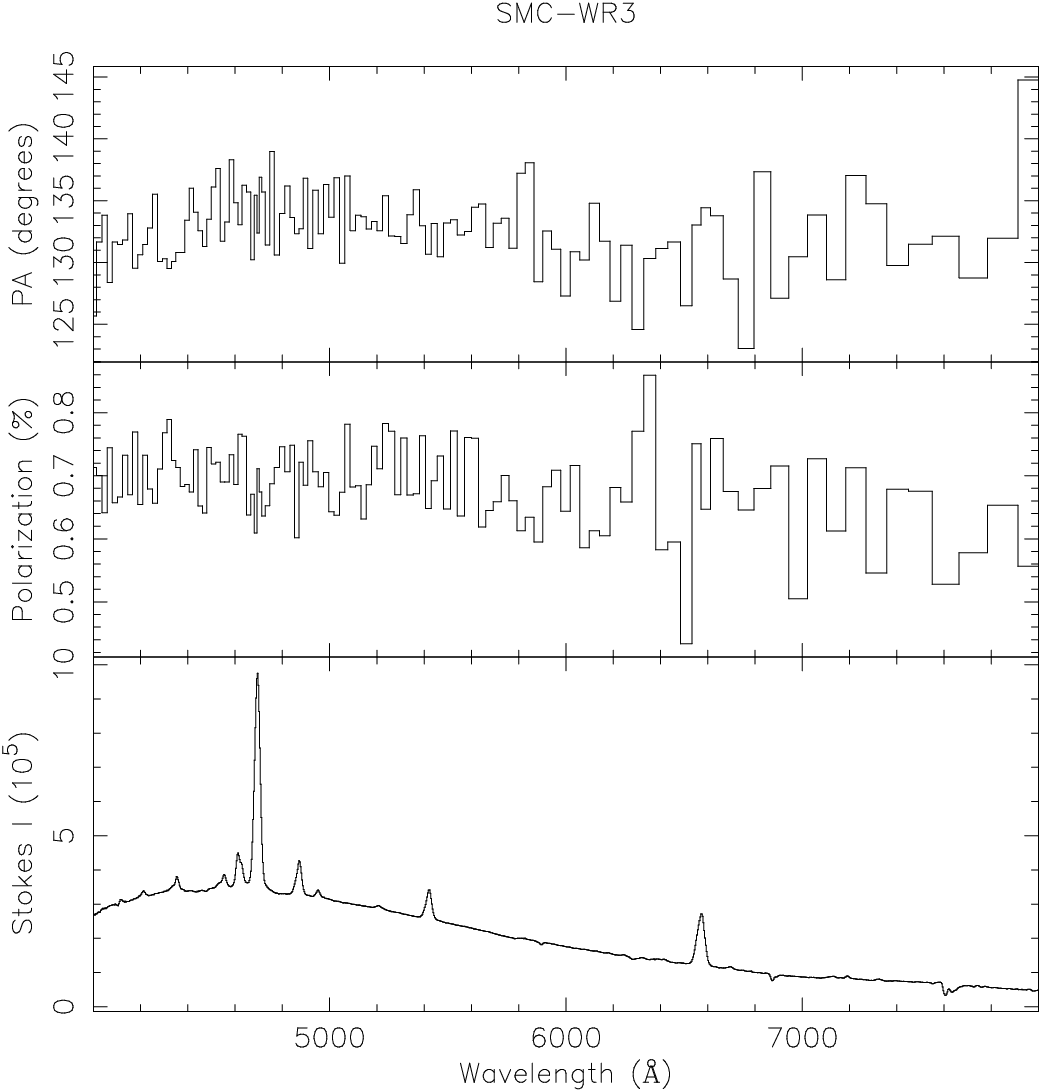}
\epsfxsize=0.25\textwidth\epsfbox{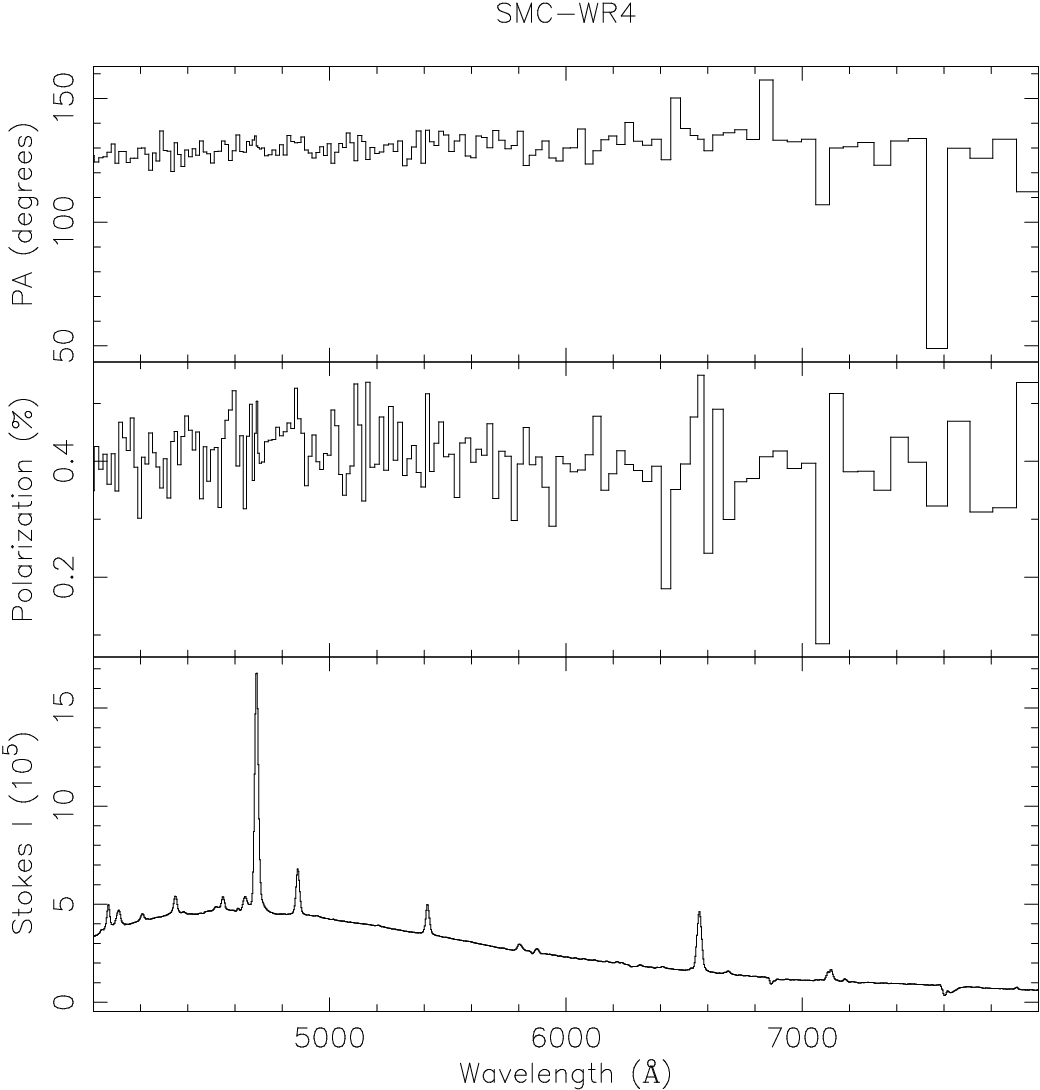}
}
\mbox{
\epsfxsize=0.25\textwidth\epsfbox{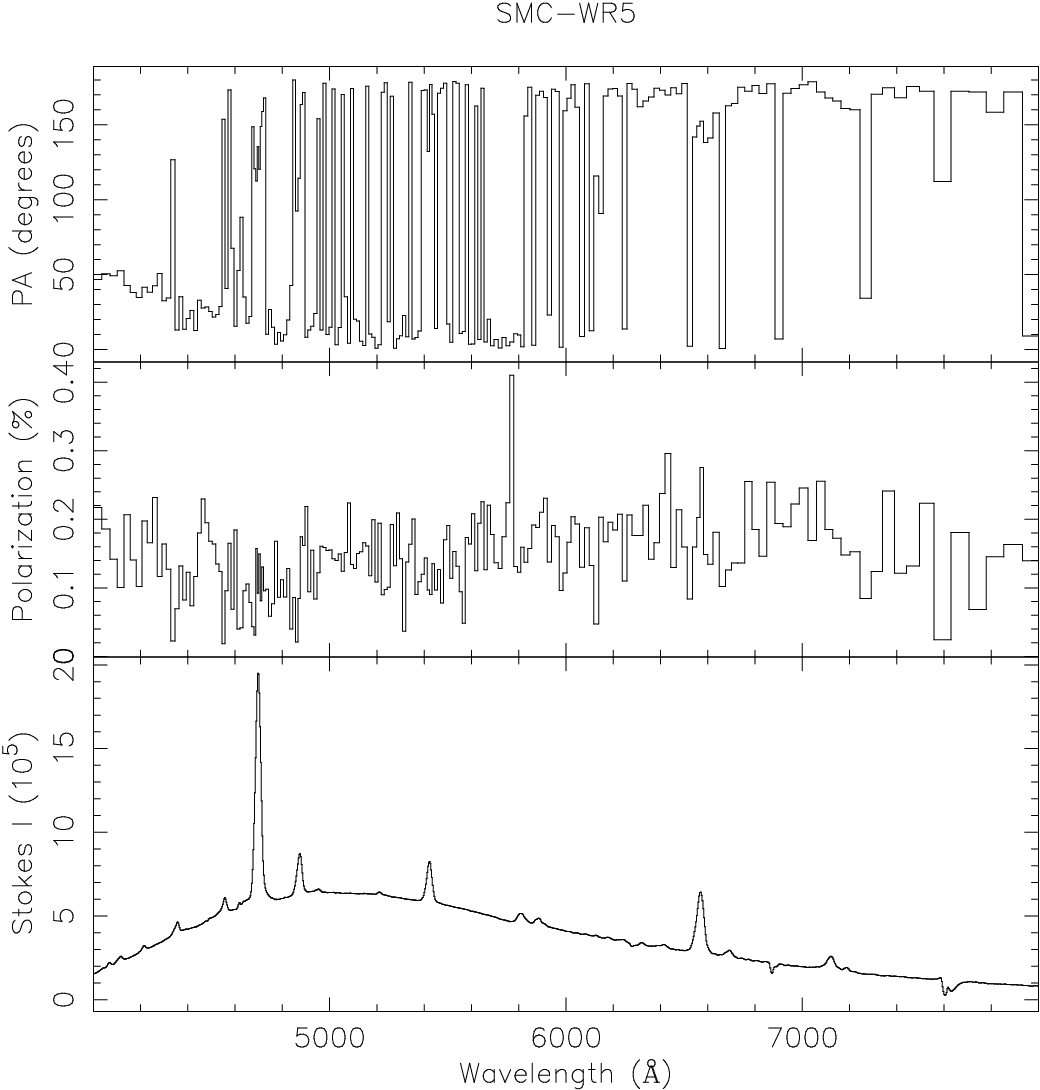}
\epsfxsize=0.25\textwidth\epsfbox{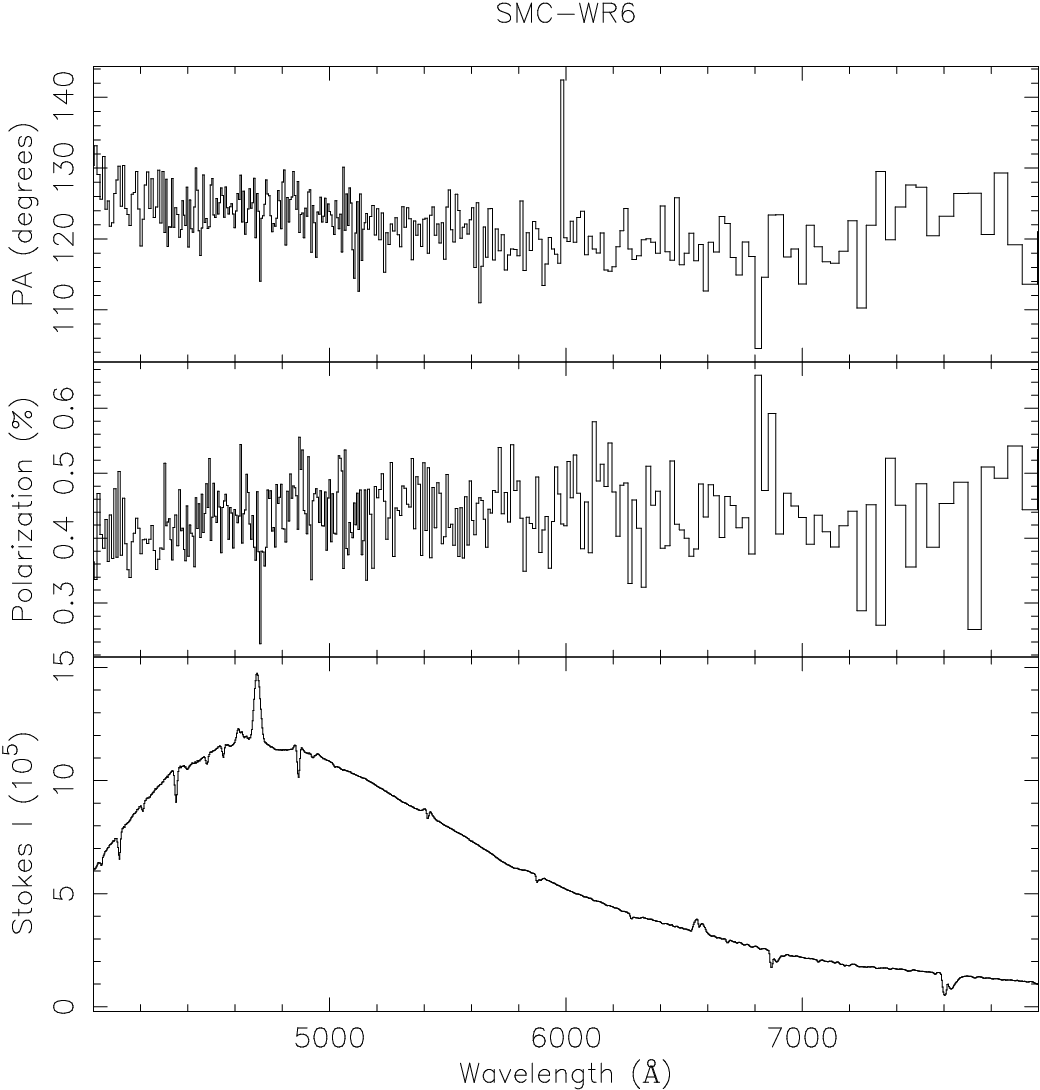}
\epsfxsize=0.25\textwidth\epsfbox{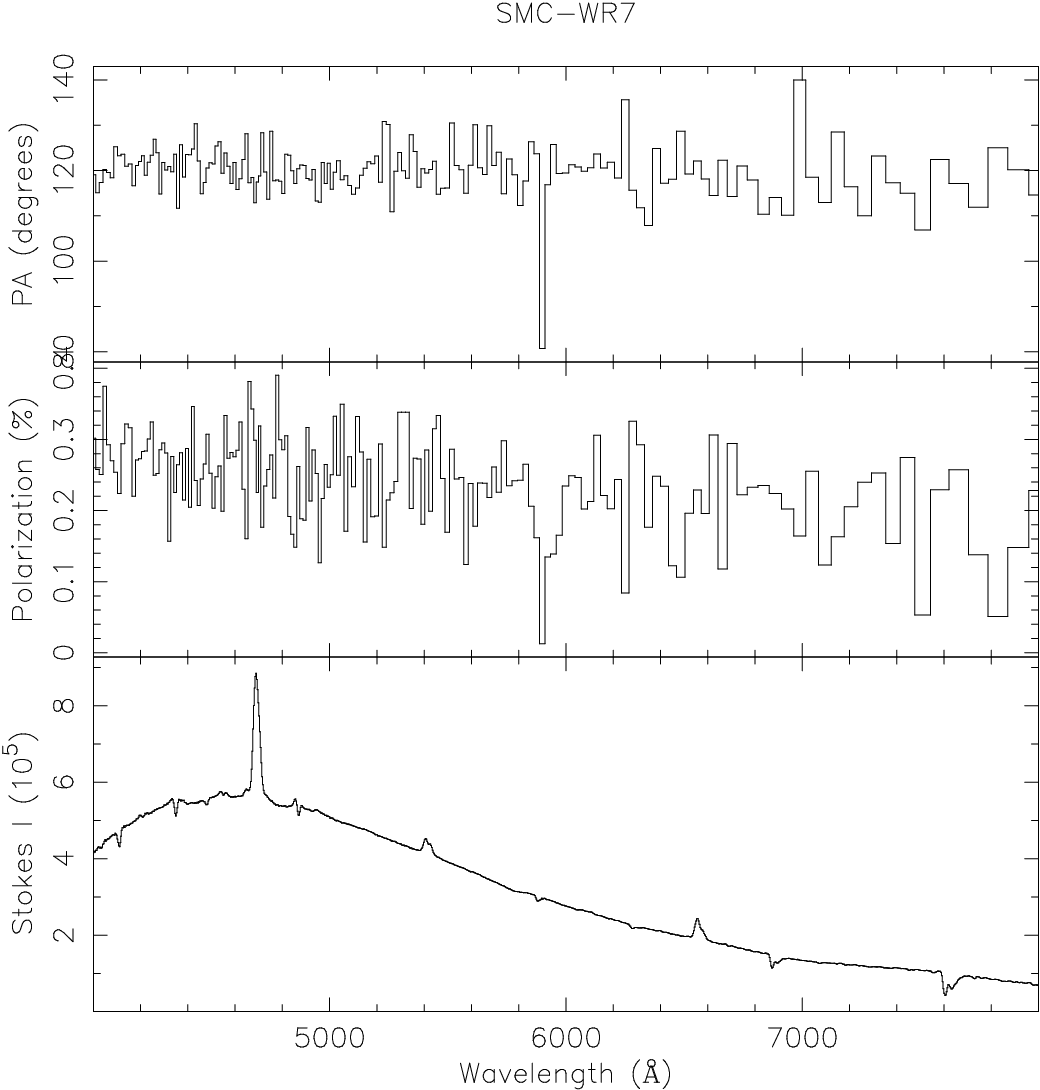}
\epsfxsize=0.25\textwidth\epsfbox{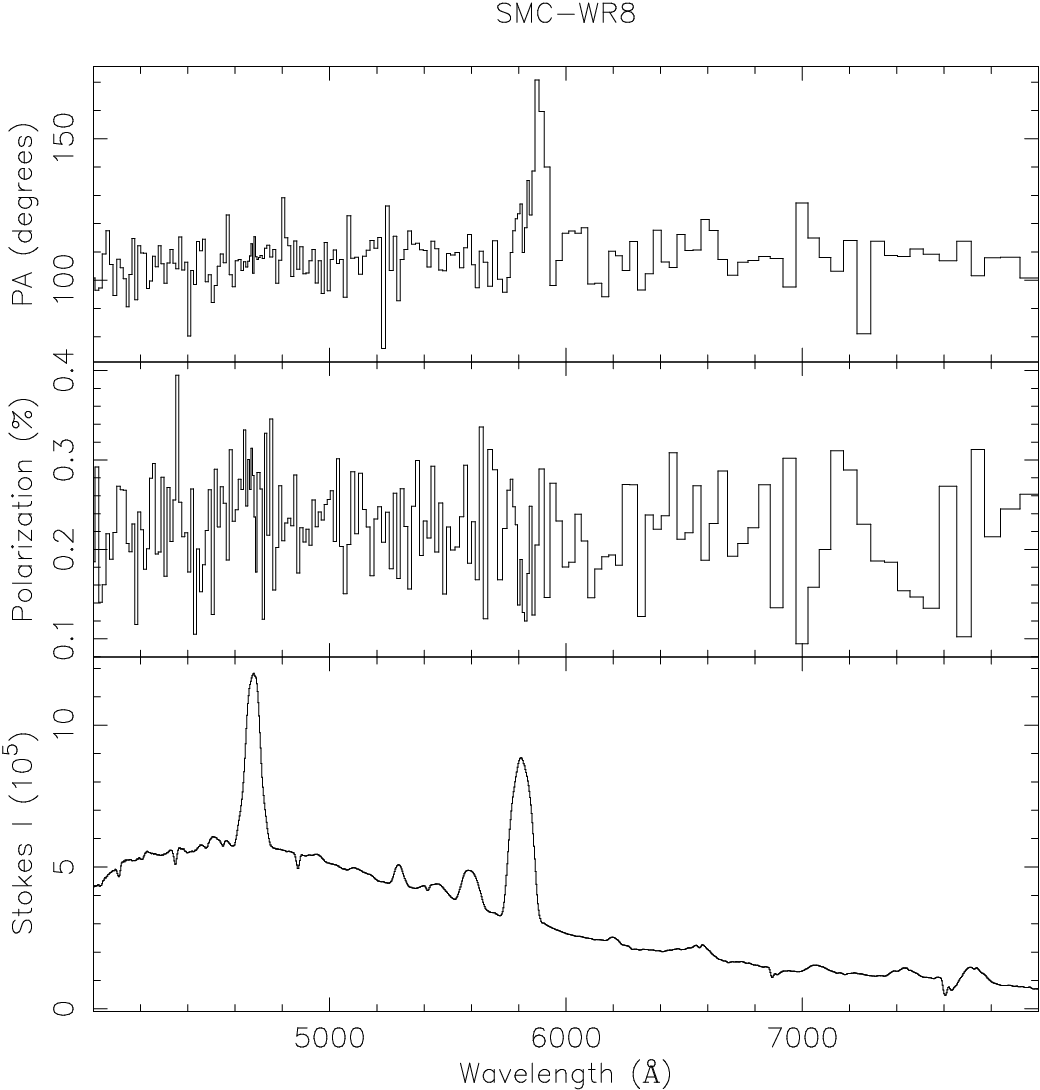}
}
\mbox{
\epsfxsize=0.25\textwidth\epsfbox{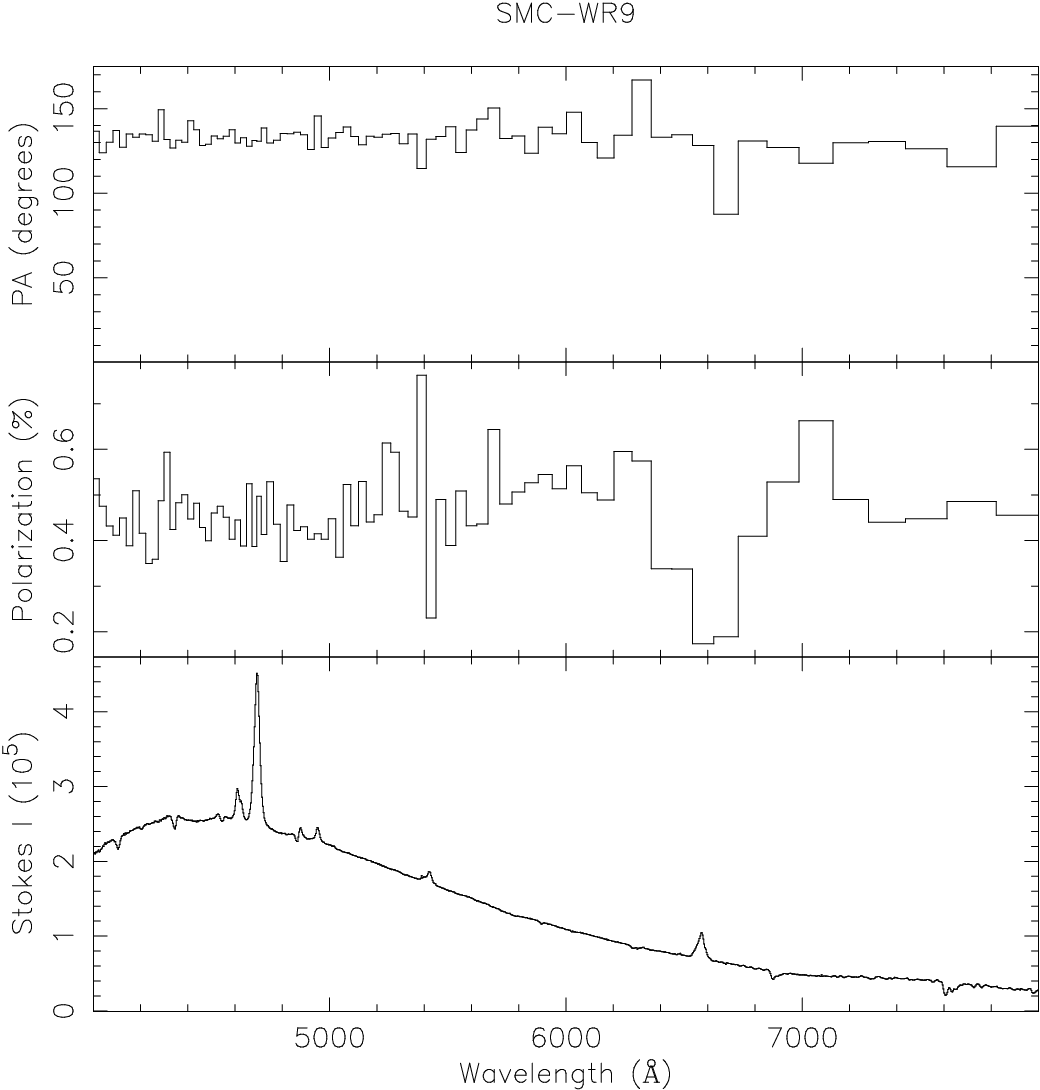}
\epsfxsize=0.25\textwidth\epsfbox{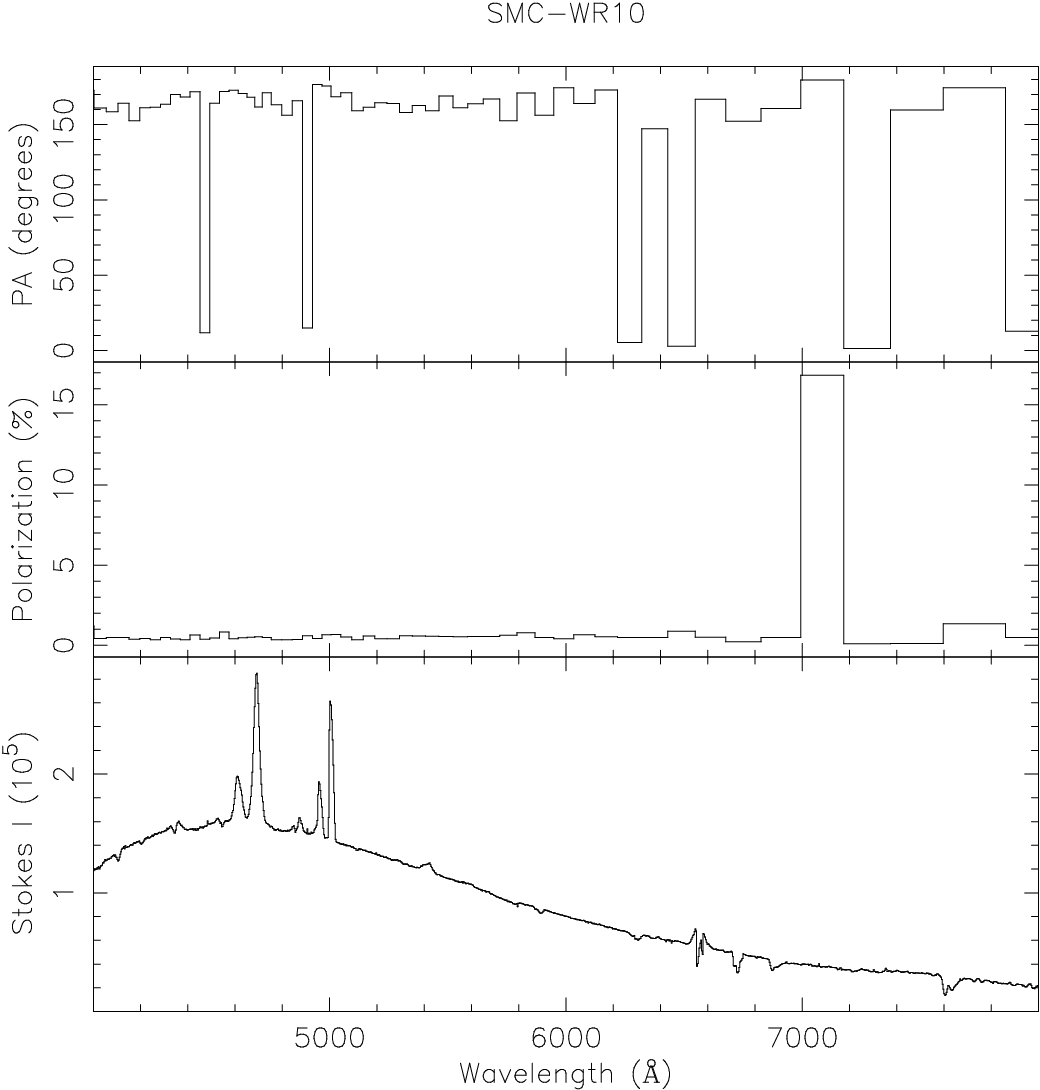}
\epsfxsize=0.25\textwidth\epsfbox{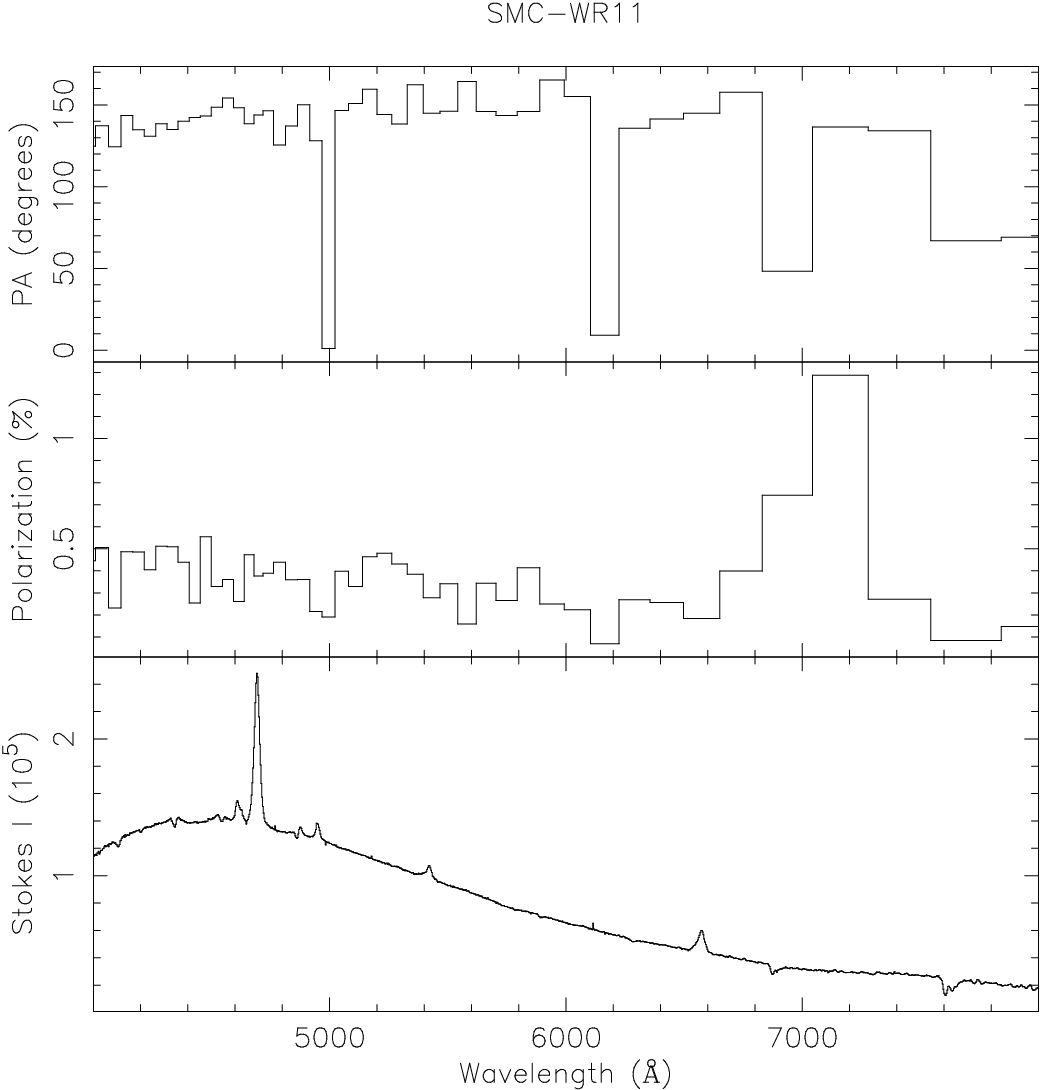}
\epsfxsize=0.25\textwidth\epsfbox{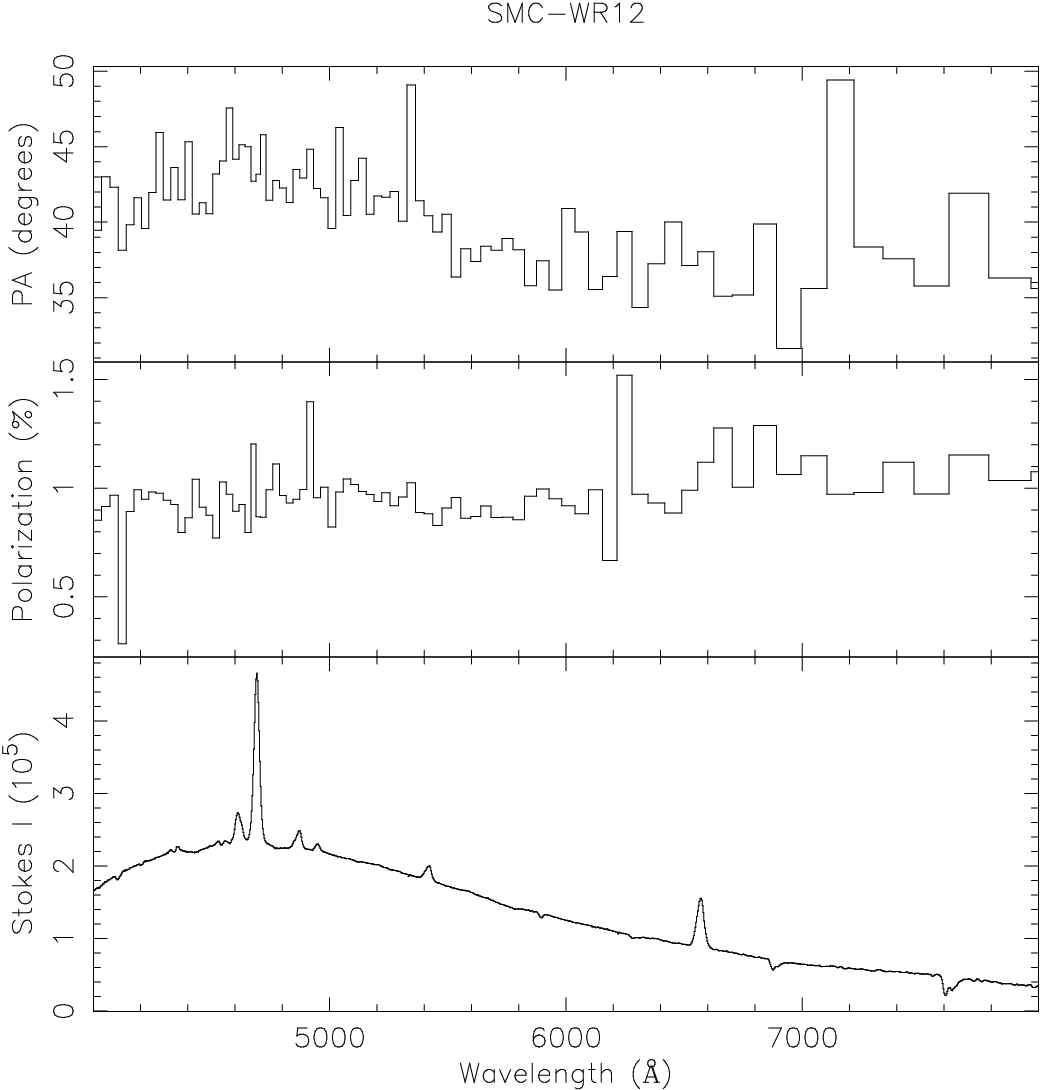}
}
\caption{Polarization spectra of the SMC WR objects.
Stokes I spectra are shown 
in the lowest panels of the triplots, 
the levels of \%Pol in the middle panel, 
whilst the PAs ($\theta$; see Eq.~2) are 
plotted in the upper panels. The data are 
rebinned such that the 1$\sigma$ error in the polarization
corresponds to 0.1\% as calculated from photon statistics.
Note that in the lowest panels, the PA is sometimes seen
 to flip with wavelength by 90-degrees, due to small changes 
between $QU$ quadrants when the observed polarization level
is close to zero.}
\label{f_smc}
\end{figure*}

The linear continuum polarization of hot stars is thought to be caused by  
scattering of stellar photons off electrons in the circumstellar environment, but 
in addition there may be an interstellar component to the measured level of polarization.  
The continuum (excluding the emission lines) polarization data for all our targets 
are summarized in Table~\ref{t_cont} in the form of the 
mean $B$-band (4200-4600\AA) percentage polarization and 
its position angle $\theta$ (columns 5 and 6). 
Polarization variability is commonplace amongst hot massive stars 
(Hayes 1975; Lupie \& Nordsieck 1987; St-Louis et al. 1987; Davies et al. 2005), and 
accordingly we do not seek perfect 
agreement between our continuum polarization measurements and those in 
earlier literature. 
In order to assess whether the data imply most  
of our targets are indeed intrinsically unpolarized, we compare our measured 
continuum polarization data to previous measurements (where available).
Given that for most objects with previous polarization data available our 
PA values are consistent with earlier measurement, it is 
safe to assume that the measured polarization for the bulk of our targets is 
of interstellar origin, with possibly a small intrinsic polarization component 
in some cases (see below). 

Plots of the spectropolarimetric data are presented in the different panels of Figs.1-4. 
The polarization spectra are
presented as triplots, consisting of Stokes I (lower panel), $P$ (middle panel), and 
the PA $\theta$ (upper panel).

\subsection{SMC objects}
\label{s_smc}

The line polarimetry data for the 12 SMC objects are shown in Fig.\ref{f_smc}. 
We do not find any evidence for line-effects in either 
the polarization percentage or the PA in our SMC data. 
Interestingly, for WR\,5, which is the famous WR/LBV HD\,5980, we have 
additional data available from our earlier P78 run (over a slightly different wavelength range). 
This P\,78 data is plotted in Fig.\,2. Here, firm line effects are detected 
across several of the strong He {\sc ii} emission lines. 

Note that although at face value there appear
to be line effects present in the triplot of WR\,8, the PA excursions are not consistent with their Stokes I line 
emission wavelengths, and we therefore attribute these spikes to be anomalous, and we do not consider WR\,8 to be 
a line-effect star. We conclude that line effects are 
rare amongst SMC stars, and we note that the line effect in HD\,5980 might be special as this object 
has shown LBV type outbursts (Koeningsberger et al. 2010 and references therein), and line effects 
amongst LBV might be more common than amongst Wolf-Rayet stars (Davies et al. 2005).

\begin{figure}
\begin{center}
\includegraphics[width=8cm]{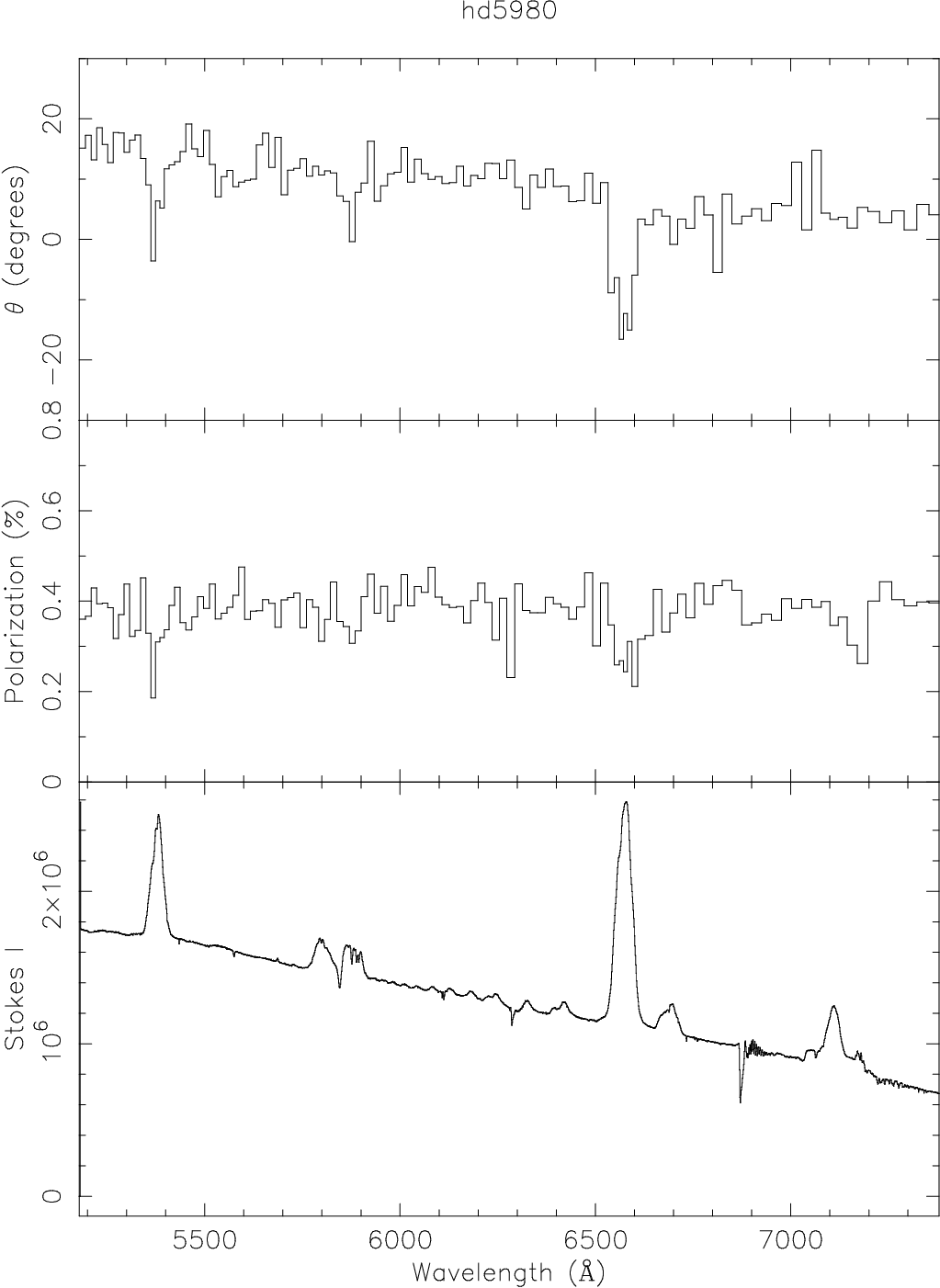} 
\end{center}
\caption{Spectropolarimetric data for the WR/LBV star HD\,5980 
taken in 2006 during the P78 run. 
The plots are in the same format as Fig.~\ref{f_smc}.}
\end{figure}

\subsection{LMC objects}
\label{s_lmc}

\begin{figure*}
\mbox{
\epsfxsize=0.25\textwidth\epsfbox{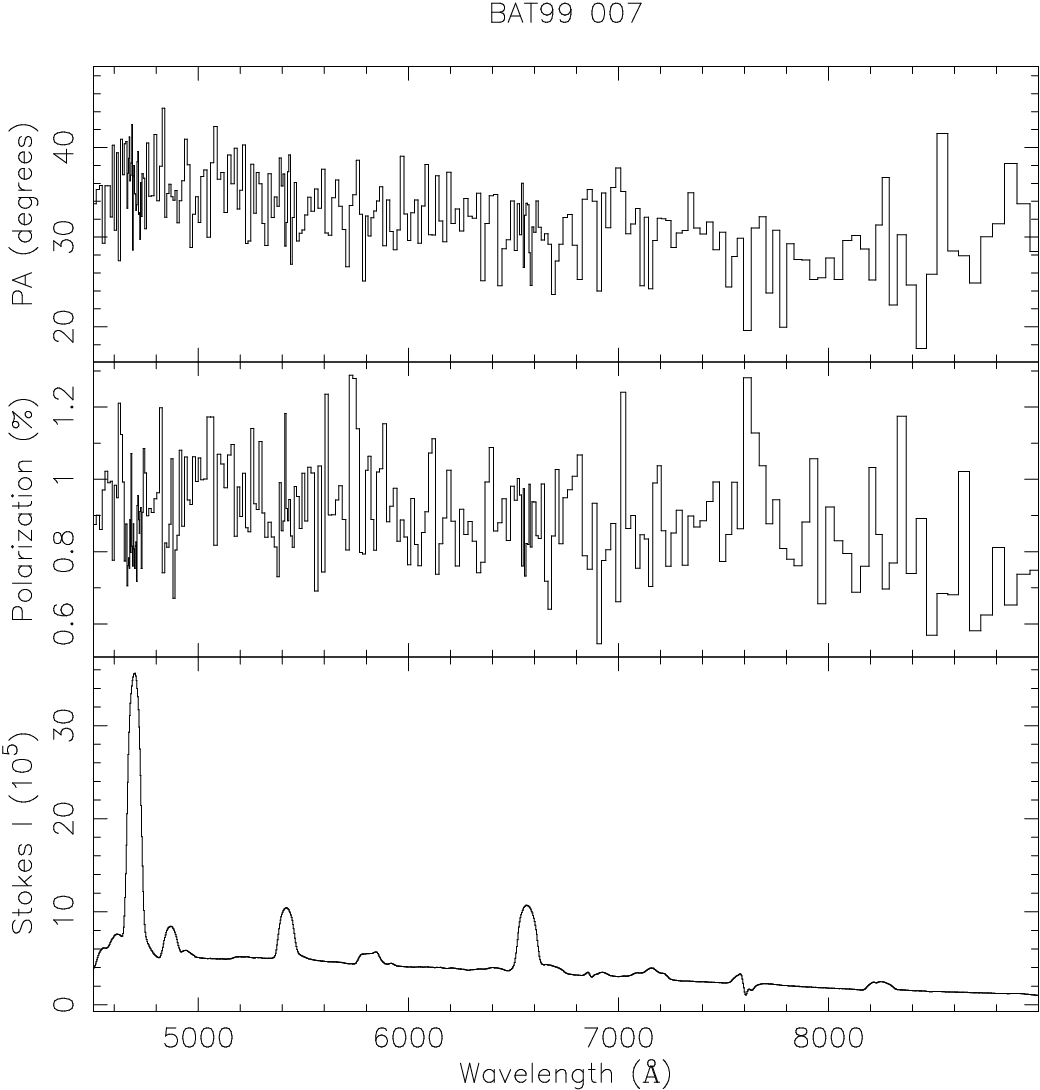}
\epsfxsize=0.25\textwidth\epsfbox{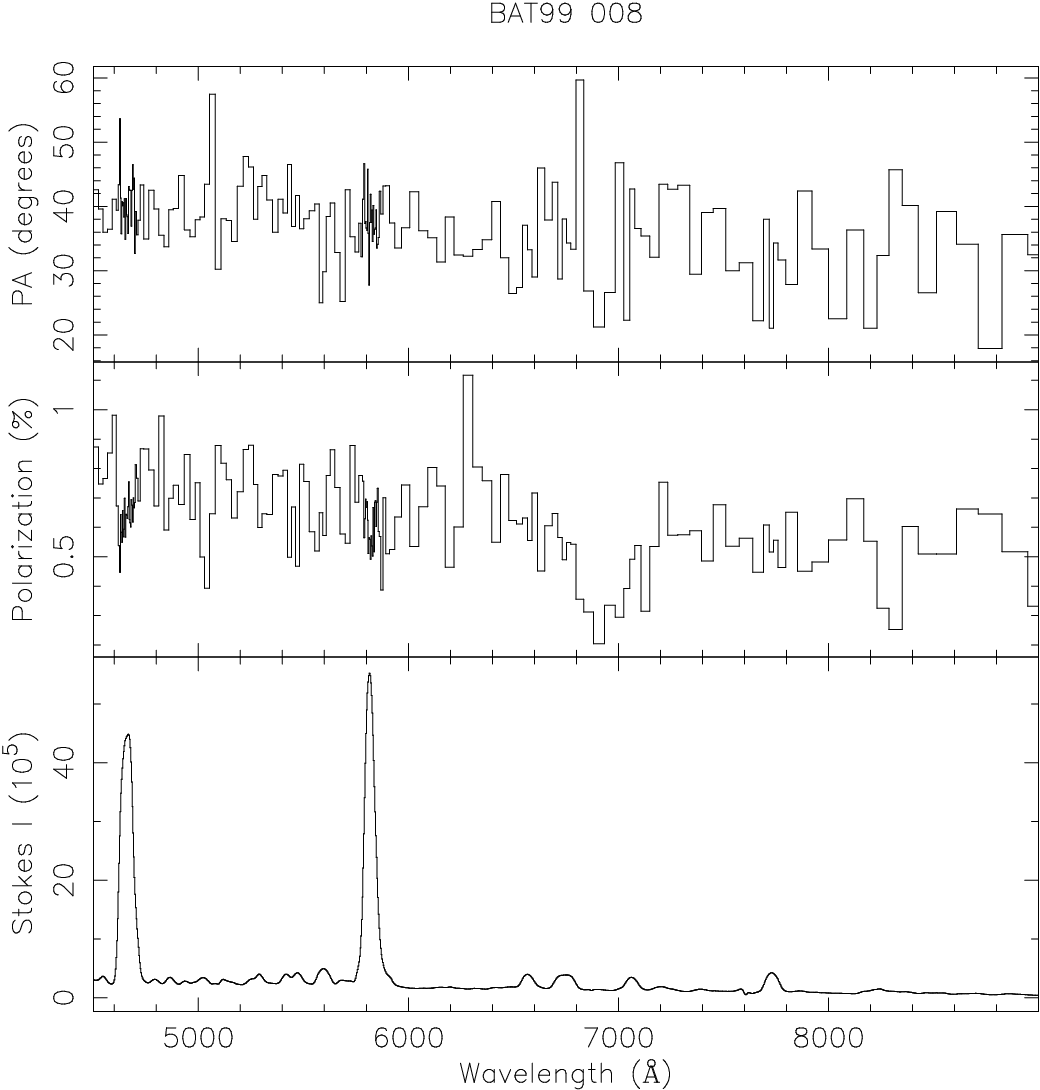}
\epsfxsize=0.25\textwidth\epsfbox{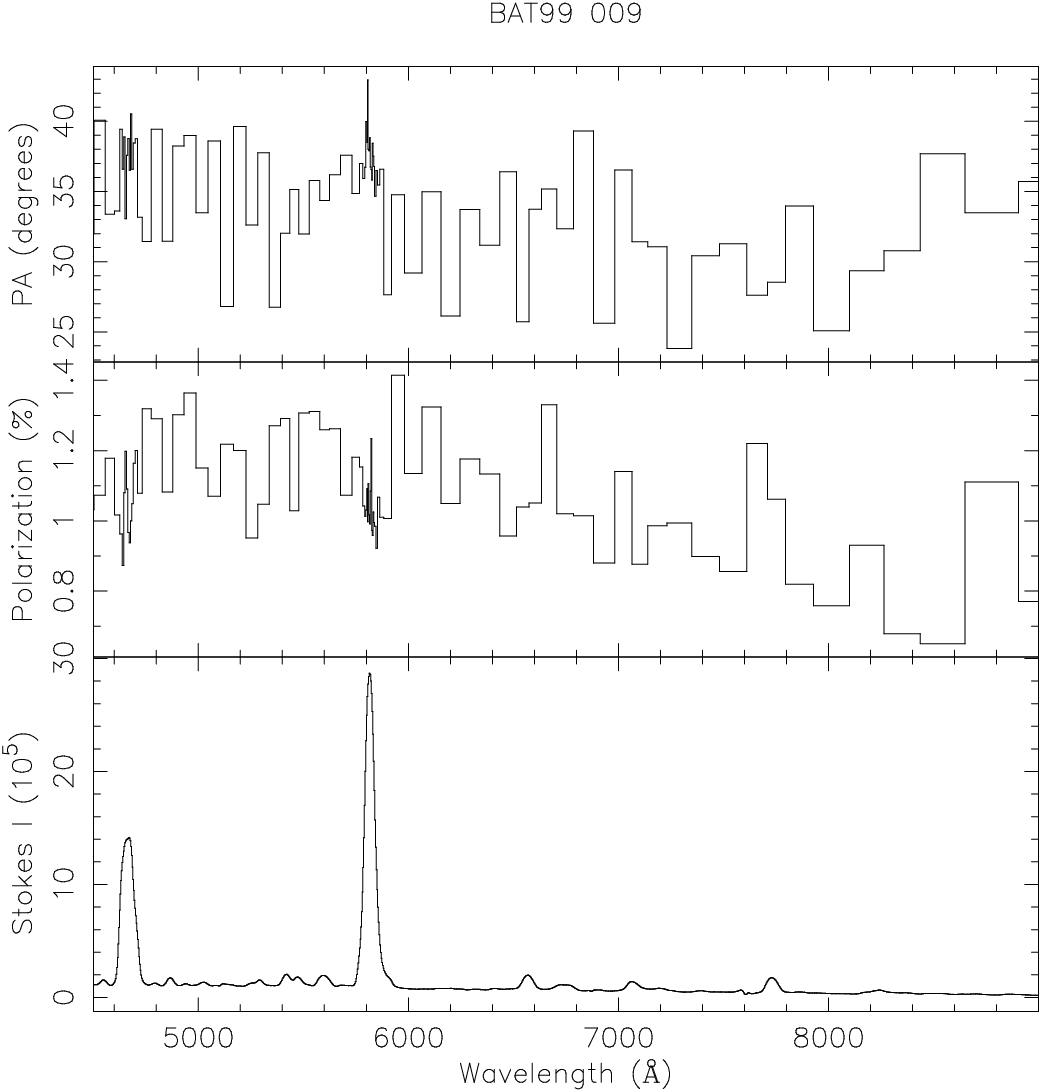}
\epsfxsize=0.25\textwidth\epsfbox{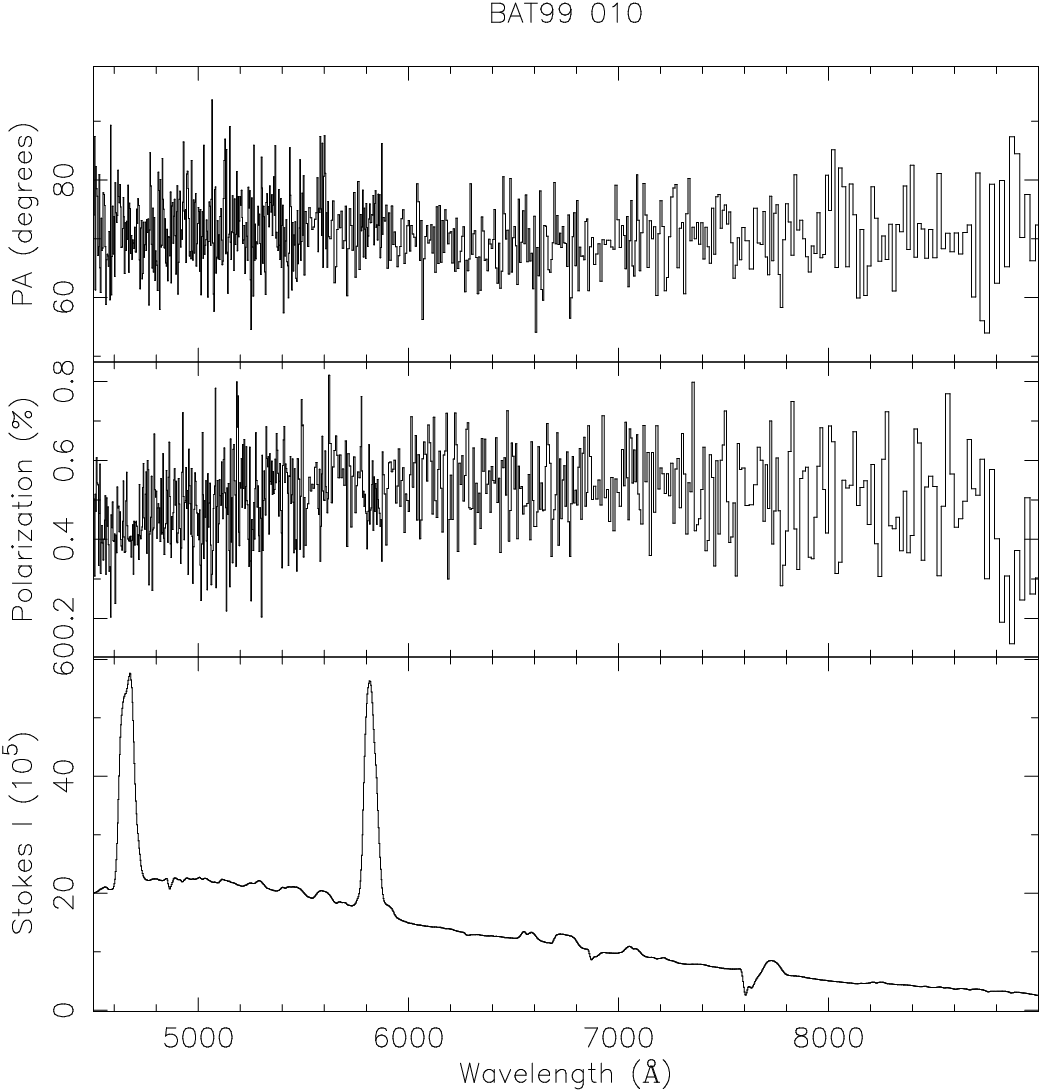}
}
\mbox{
\epsfxsize=0.25\textwidth\epsfbox{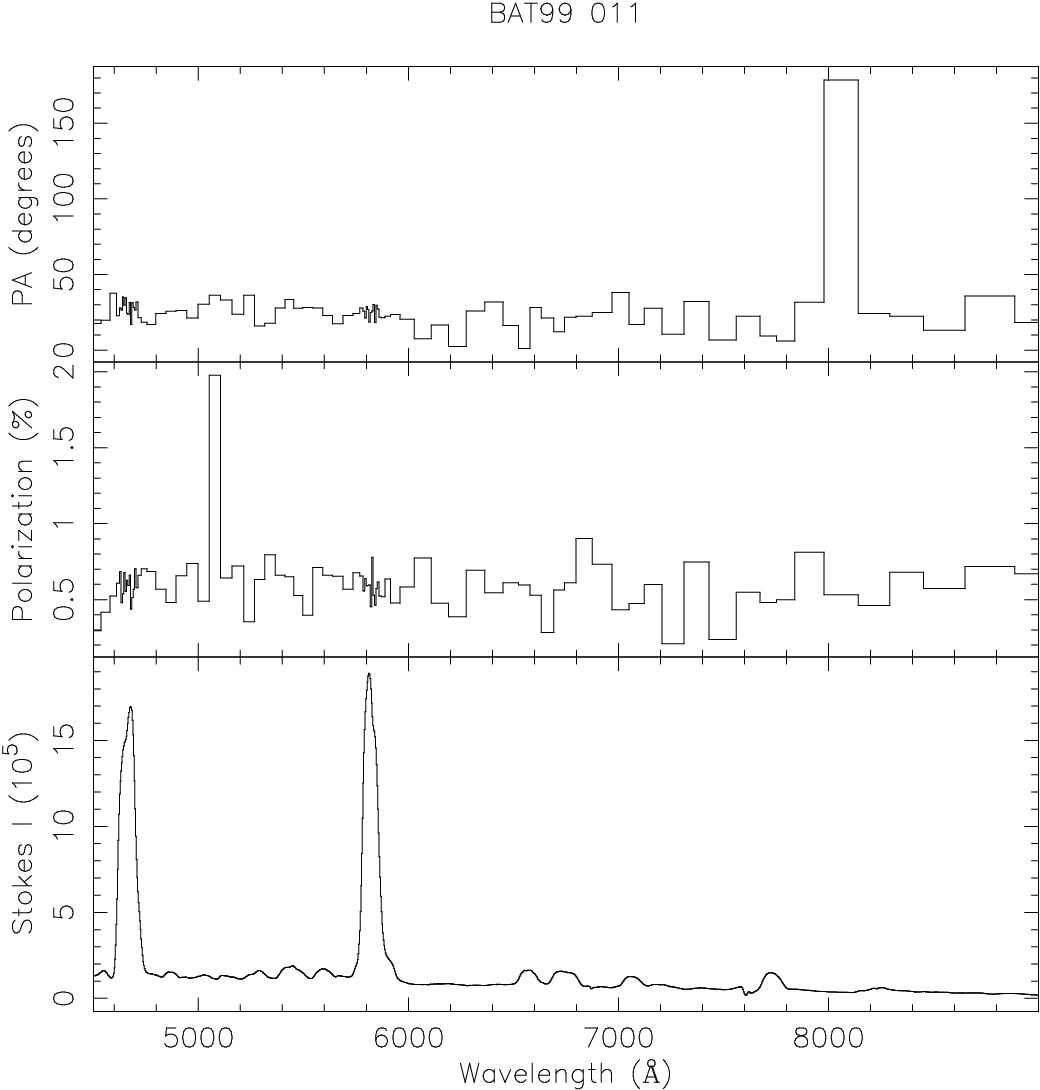}
\epsfxsize=0.25\textwidth\epsfbox{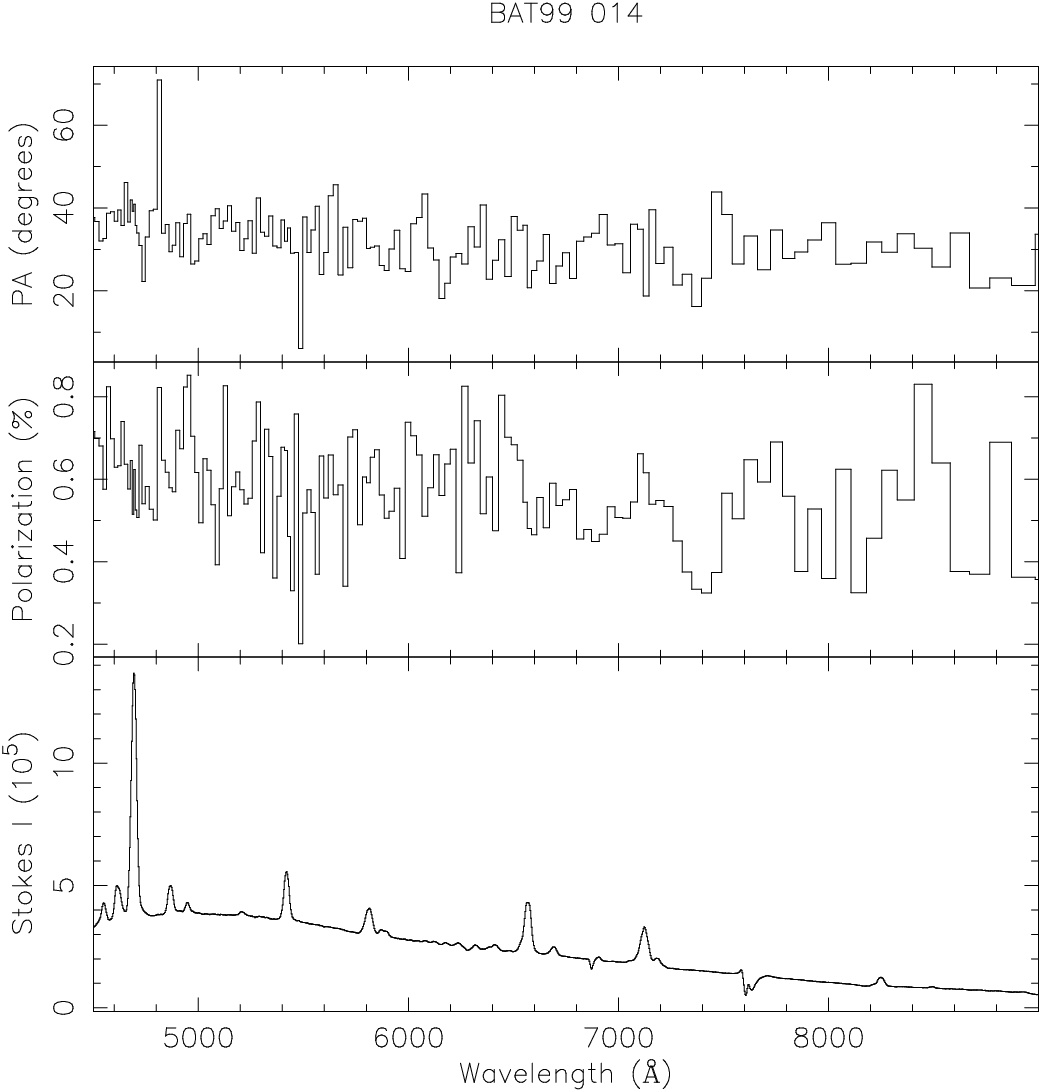}
\epsfxsize=0.25\textwidth\epsfbox{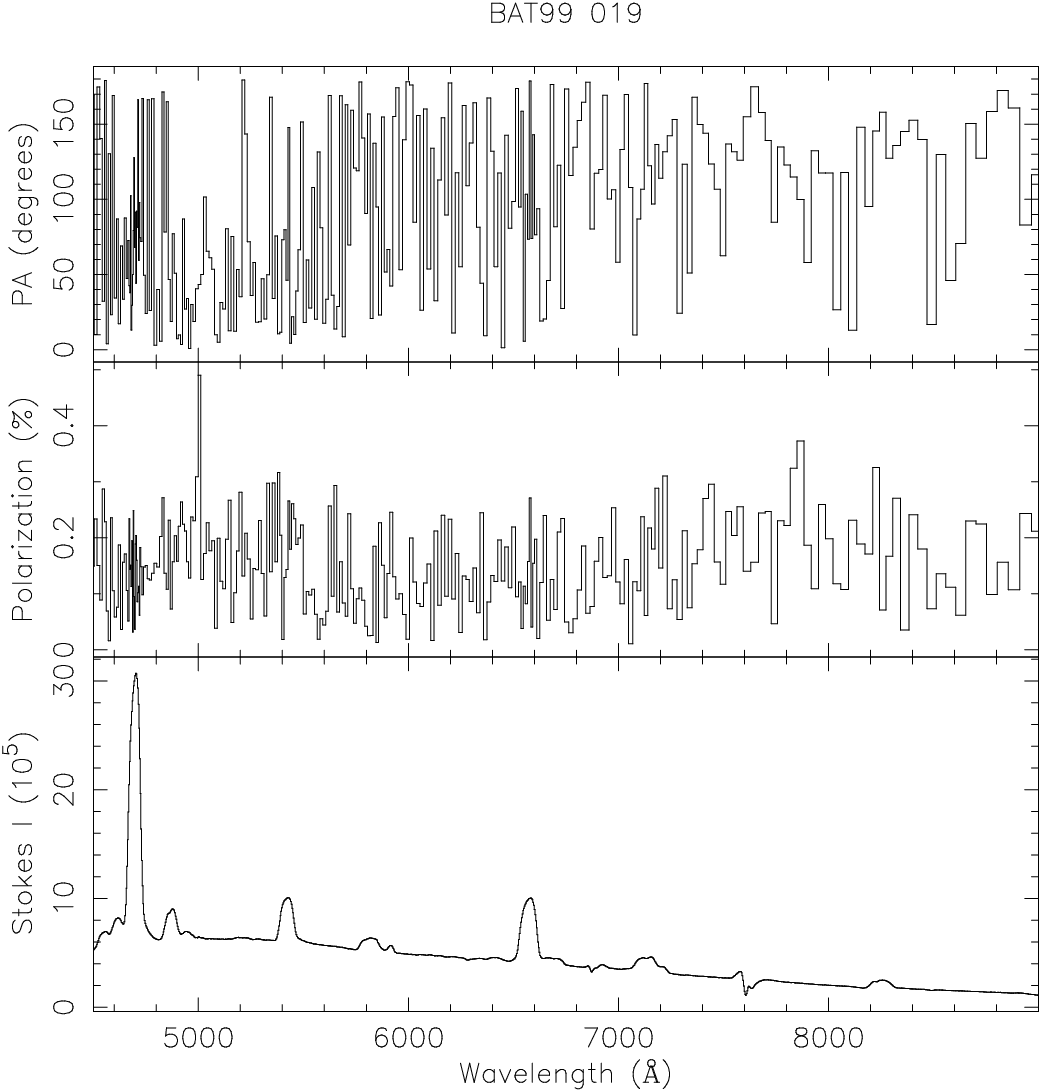}
\epsfxsize=0.25\textwidth\epsfbox{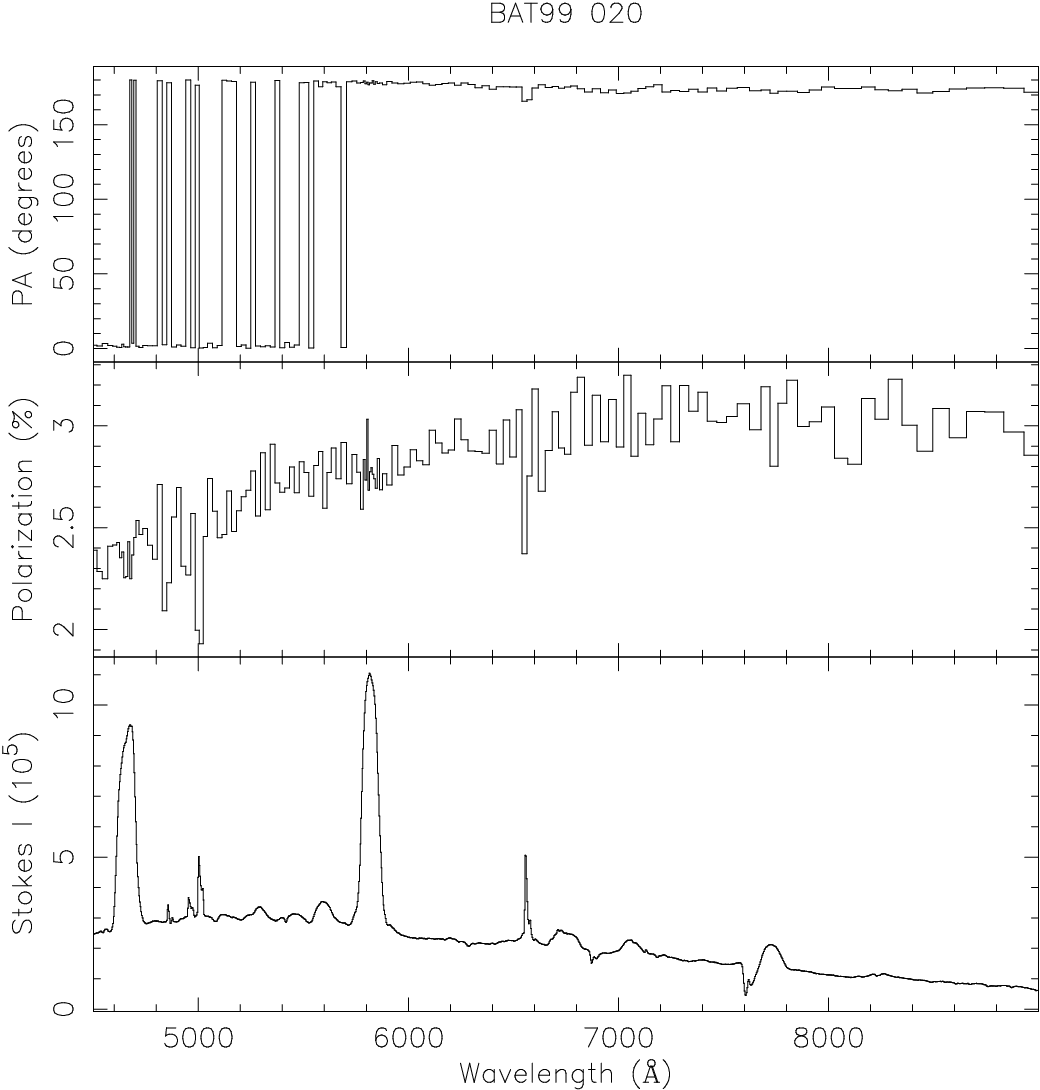}
}
\mbox{
\epsfxsize=0.25\textwidth\epsfbox{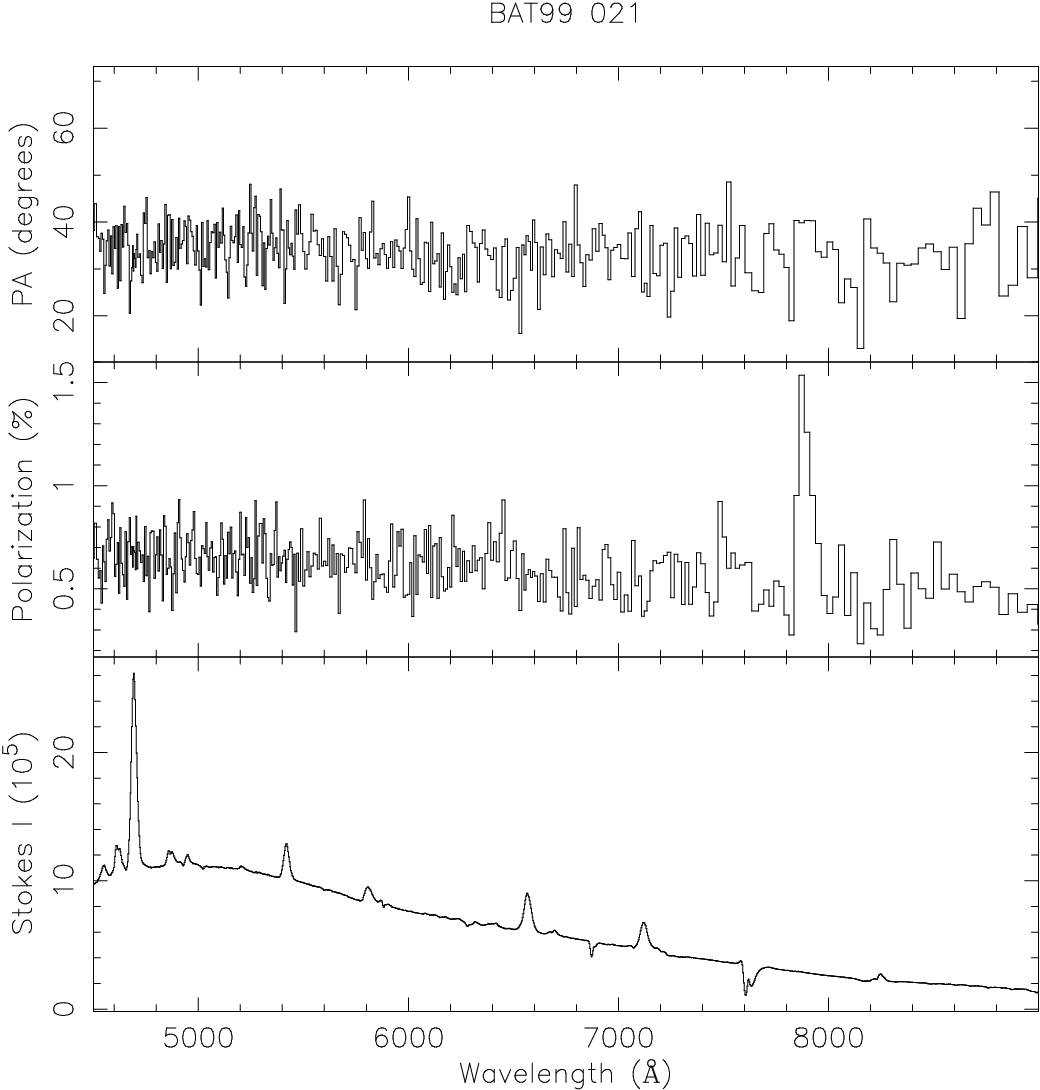}
\epsfxsize=0.25\textwidth\epsfbox{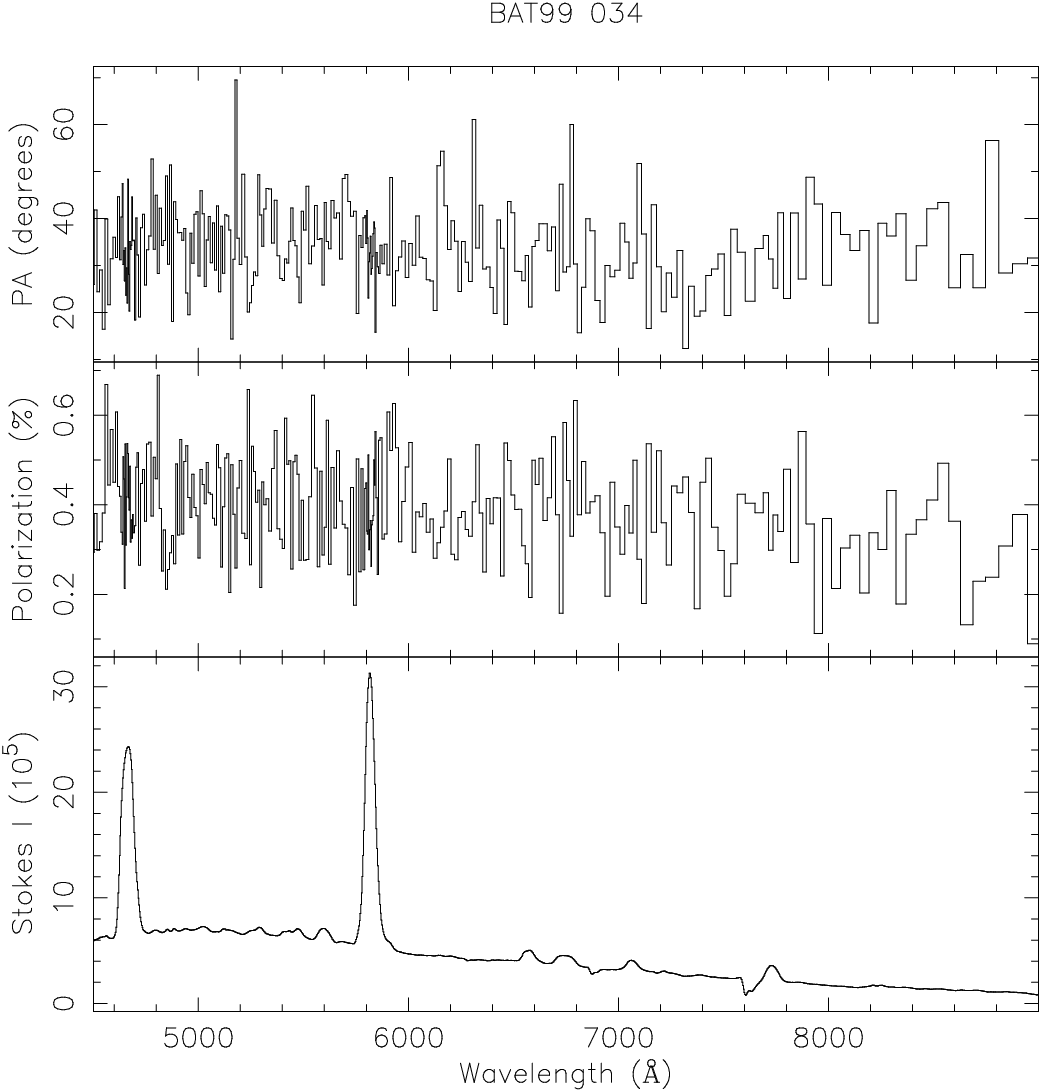}
\epsfxsize=0.25\textwidth\epsfbox{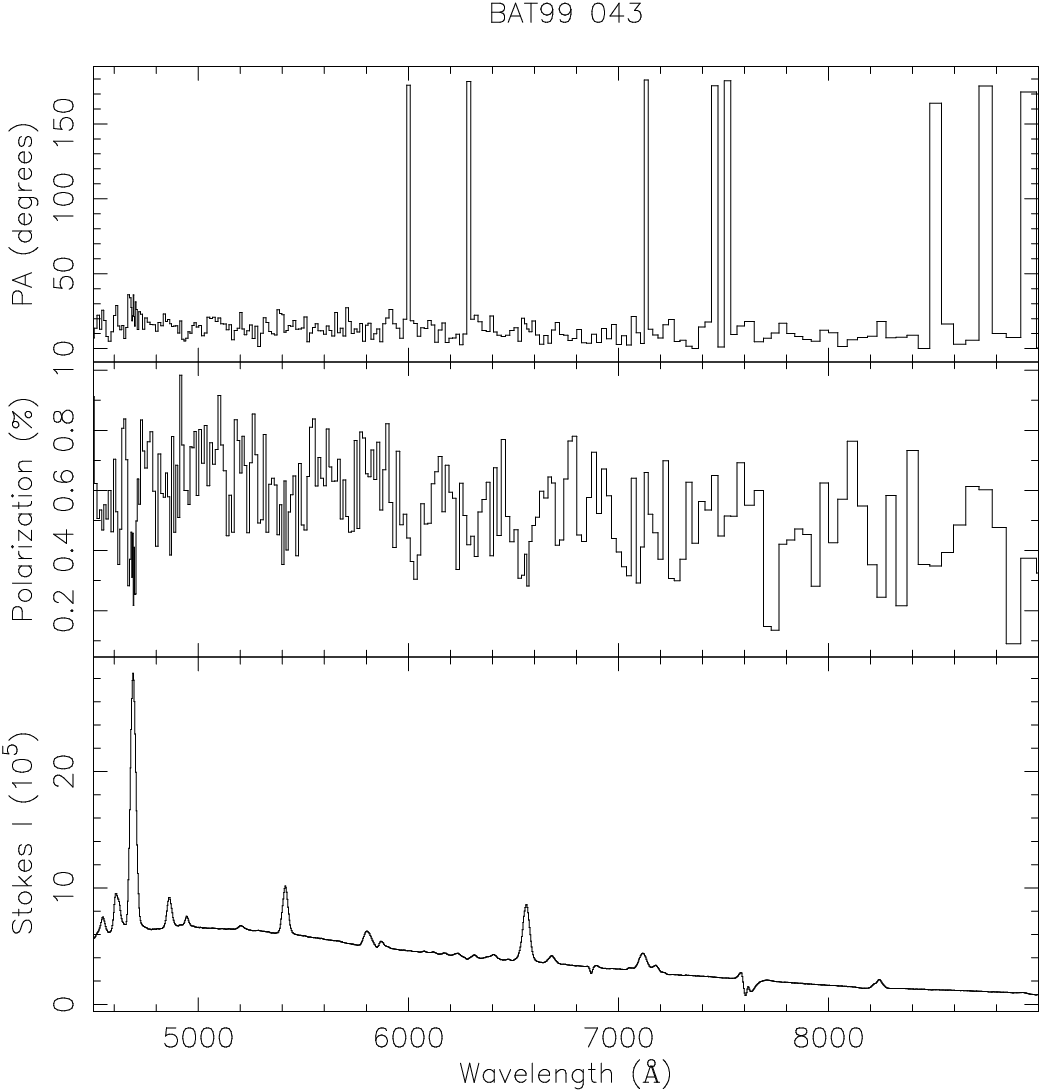}
\epsfxsize=0.25\textwidth\epsfbox{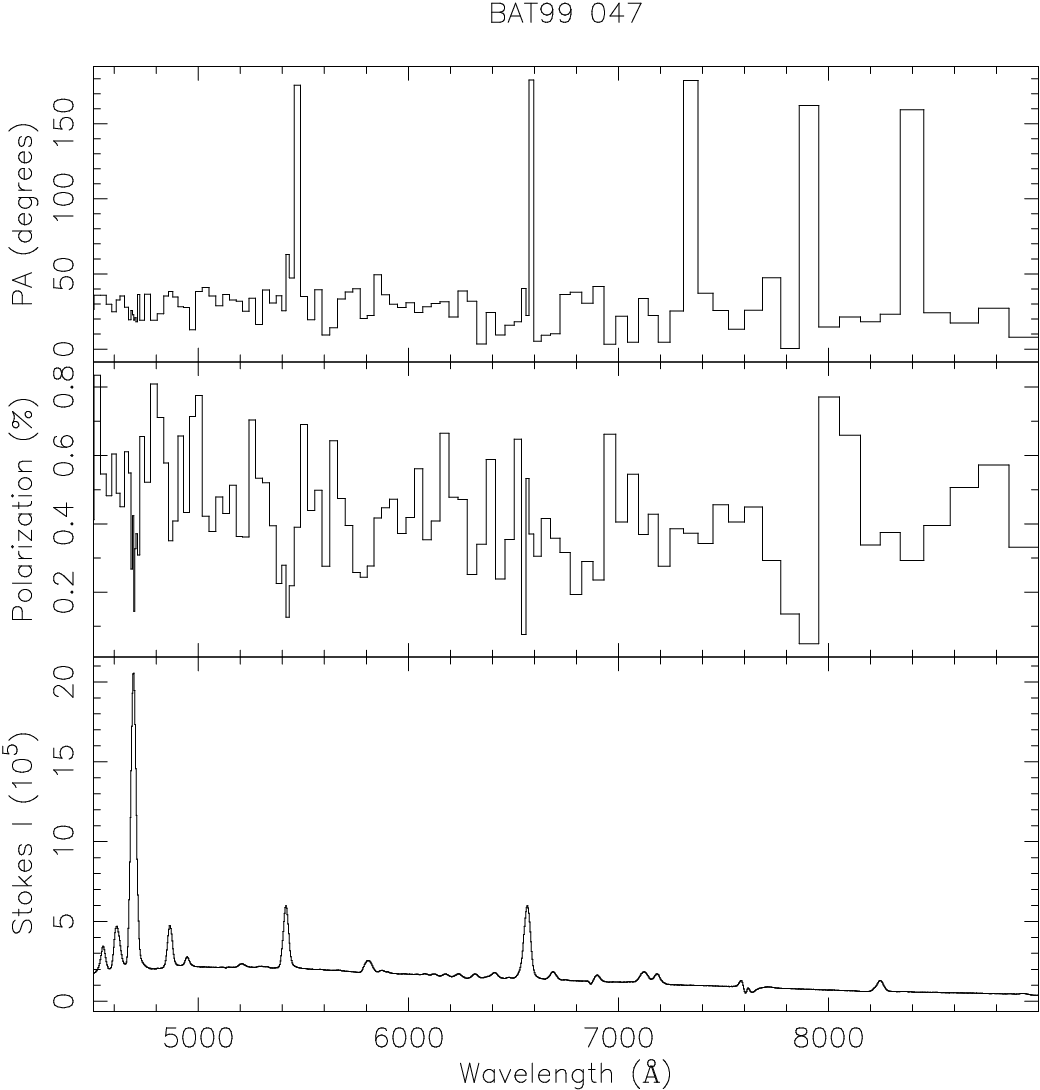}
}
\mbox{
\epsfxsize=0.25\textwidth\epsfbox{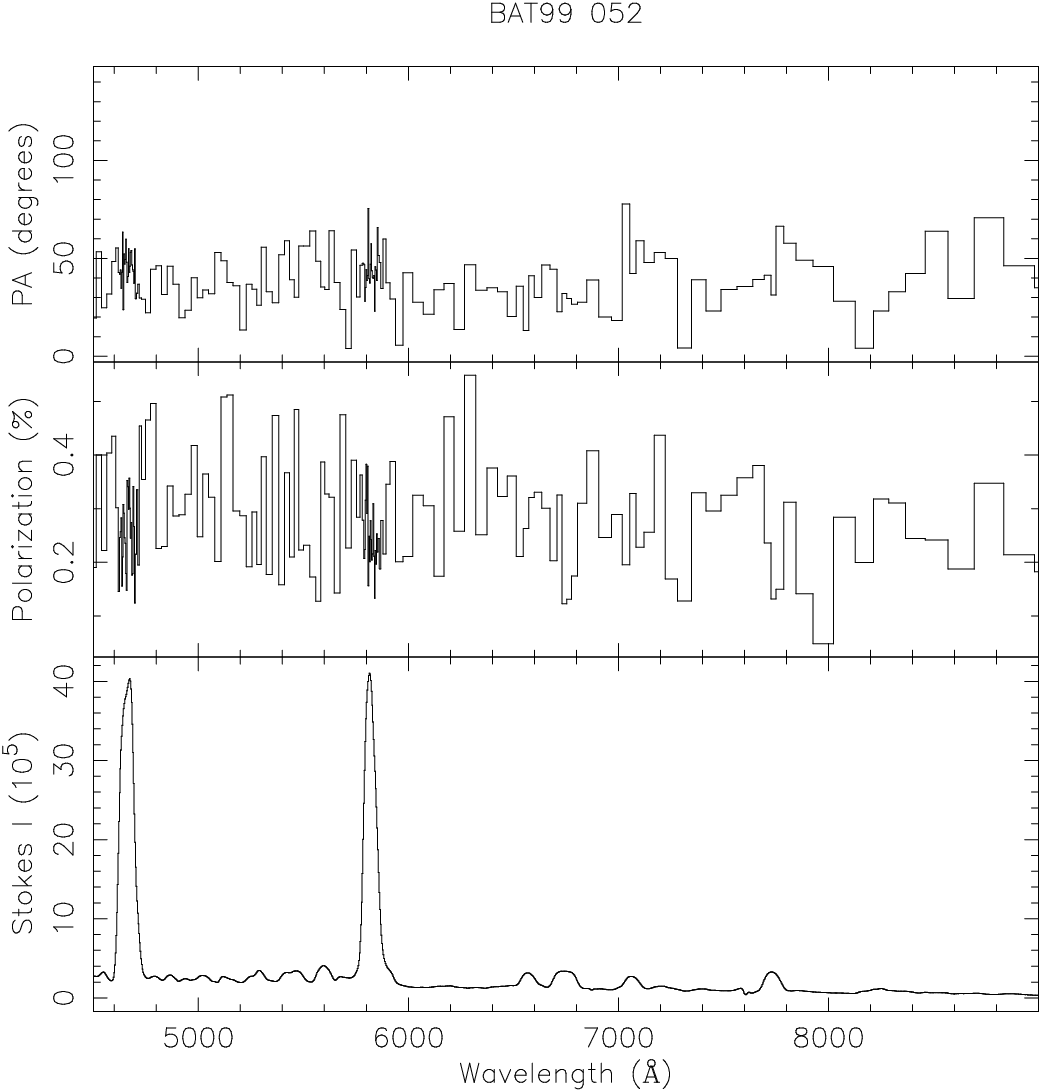}
\epsfxsize=0.25\textwidth\epsfbox{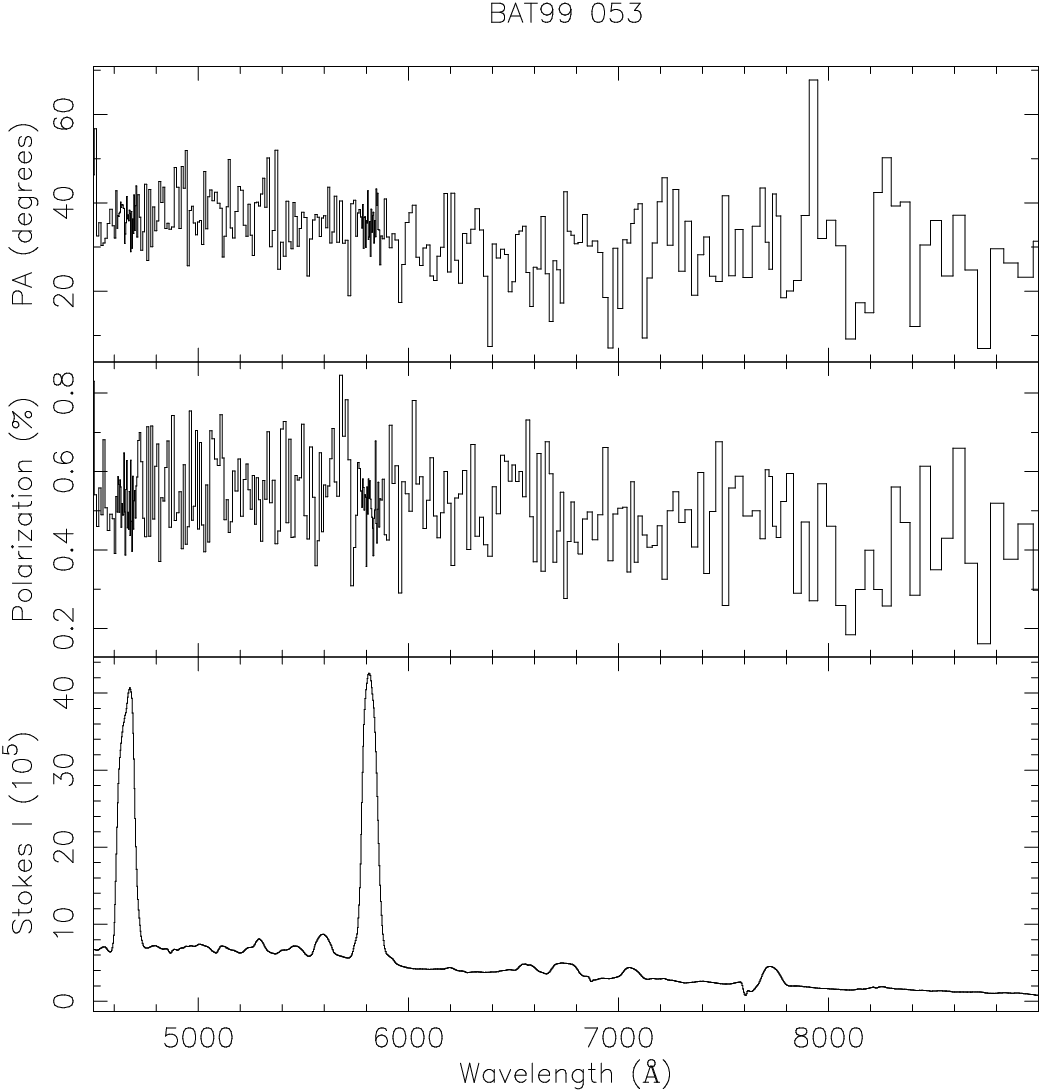}
\epsfxsize=0.25\textwidth\epsfbox{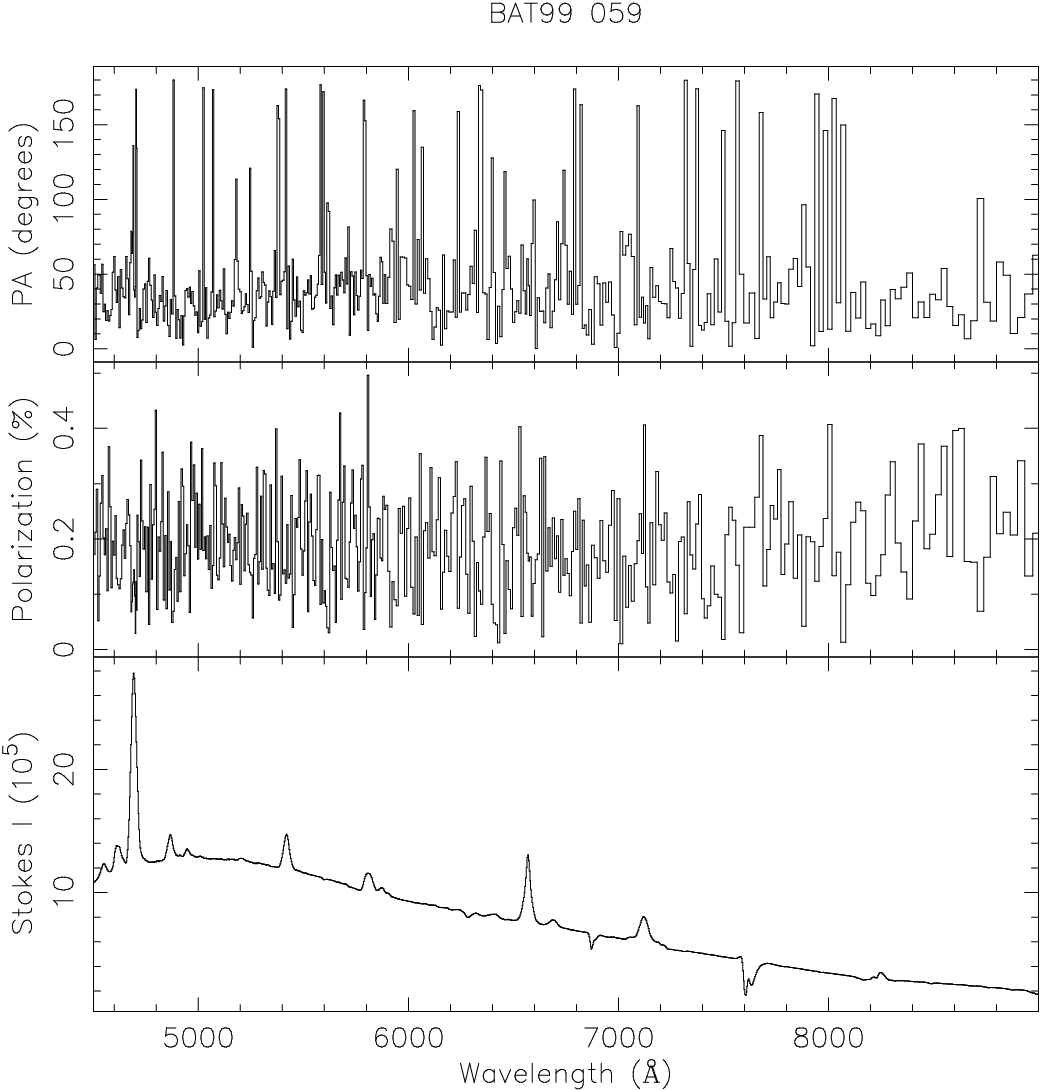}
\epsfxsize=0.25\textwidth\epsfbox{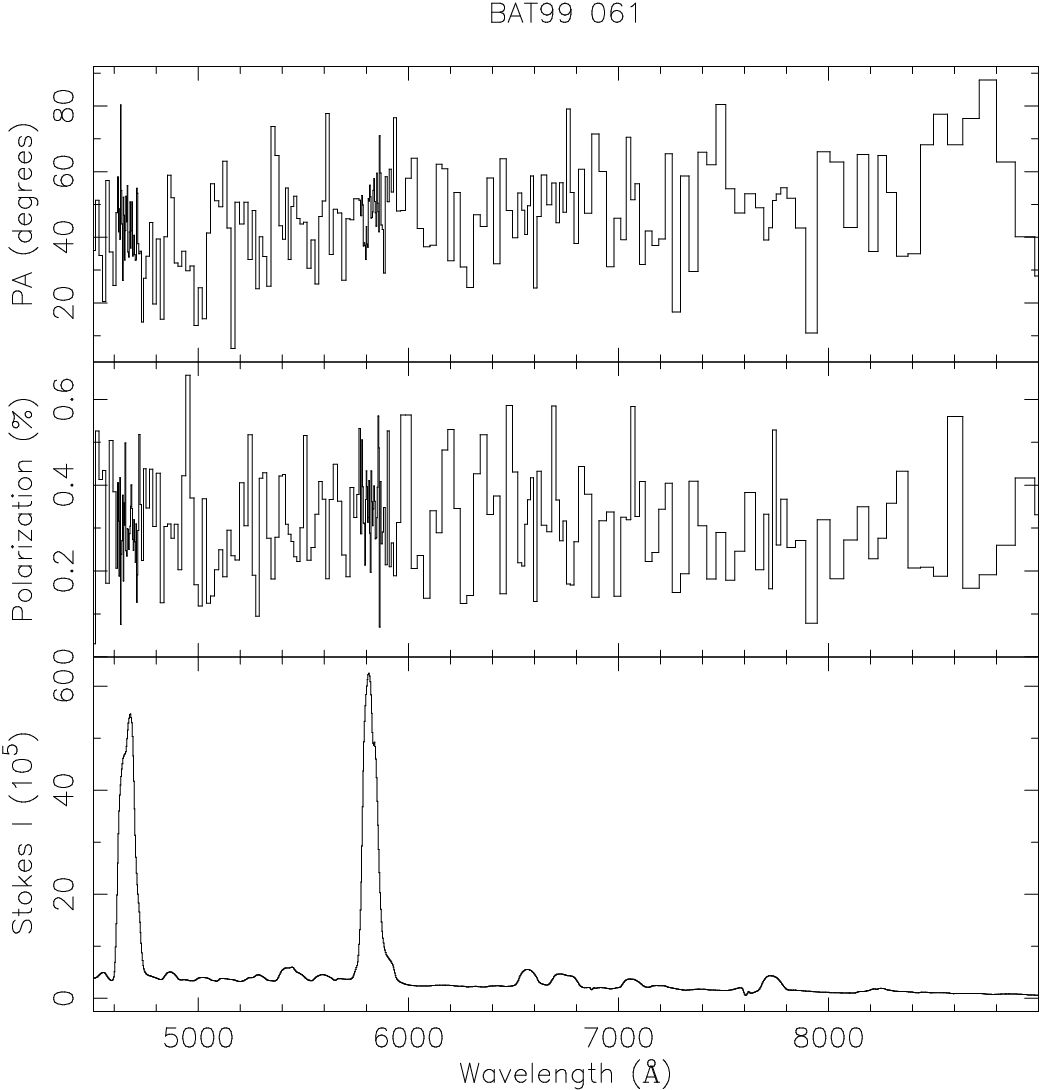}
}
\mbox{
\epsfxsize=0.25\textwidth\epsfbox{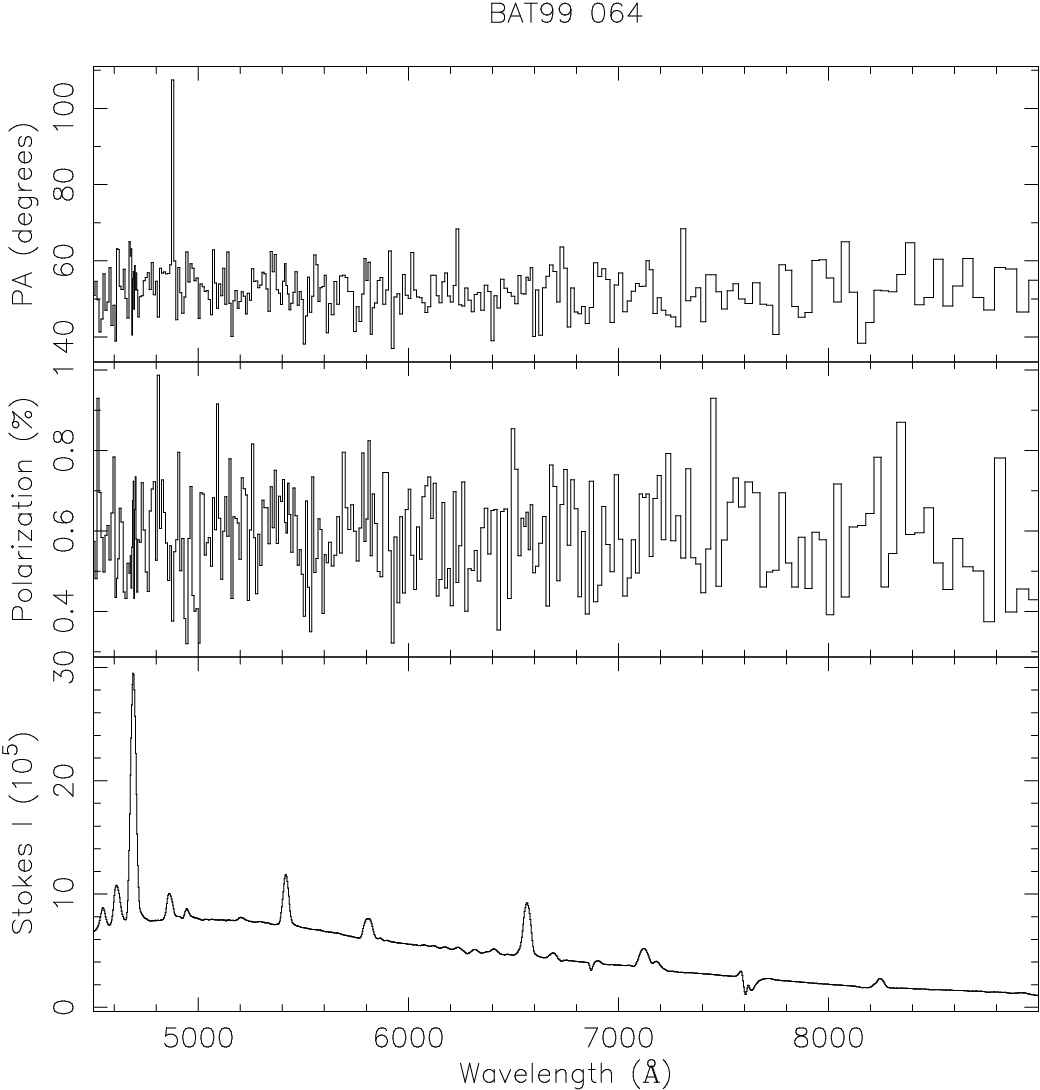}
\epsfxsize=0.25\textwidth\epsfbox{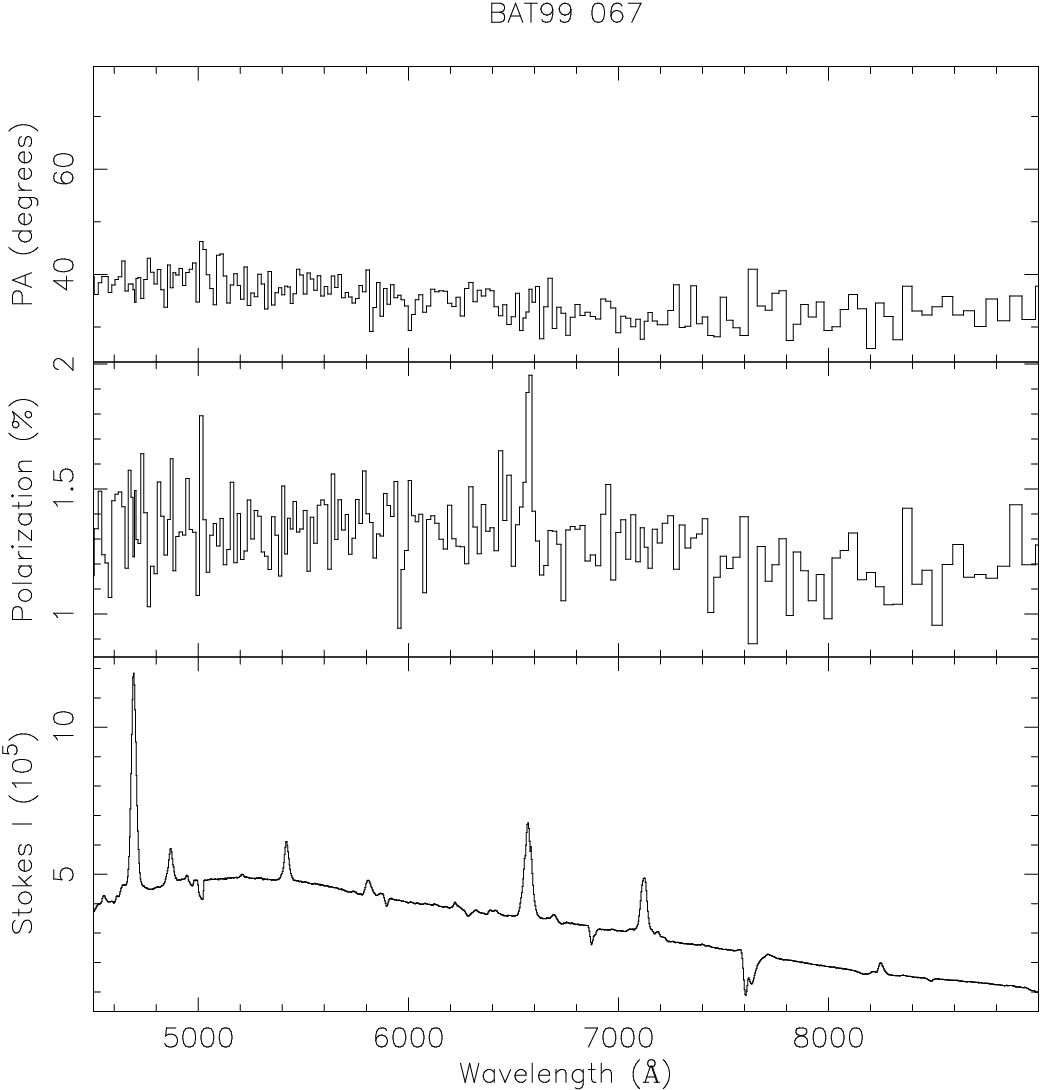}
\epsfxsize=0.25\textwidth\epsfbox{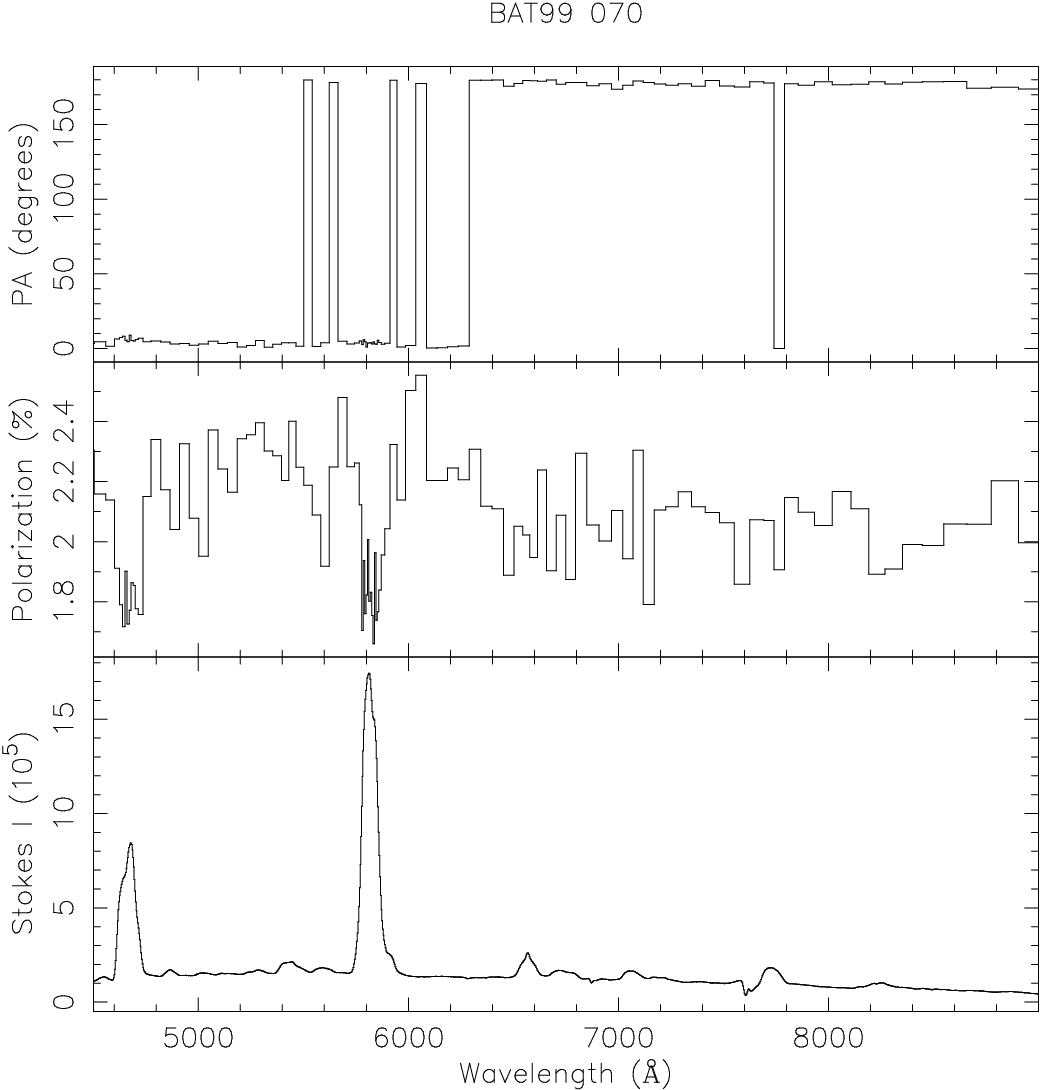}
\epsfxsize=0.25\textwidth\epsfbox{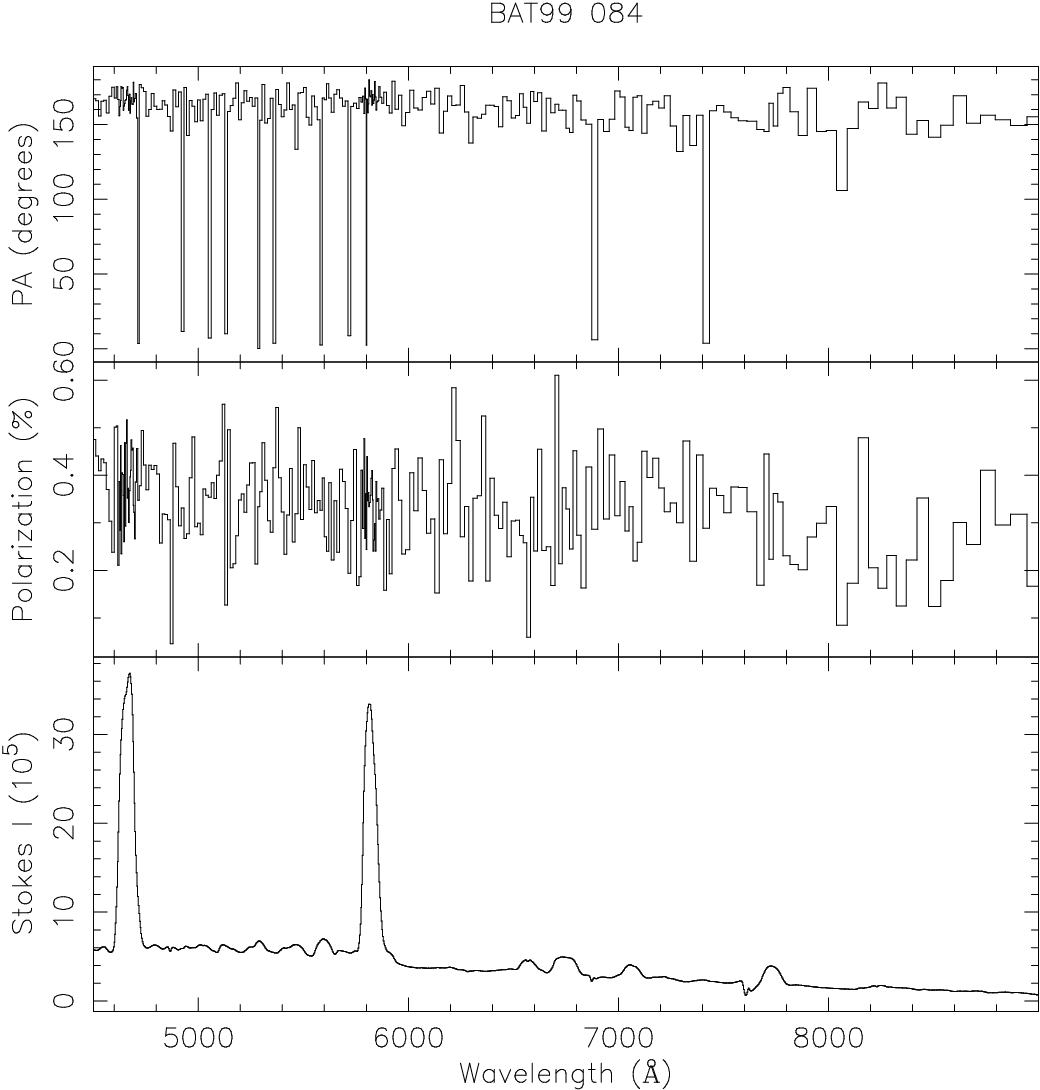}
}
\caption{Polarization spectra of LMC WR stars.
\label{f_lmc}}
\end{figure*}

\begin{figure*}
\mbox{
\epsfxsize=0.25\textwidth\epsfbox{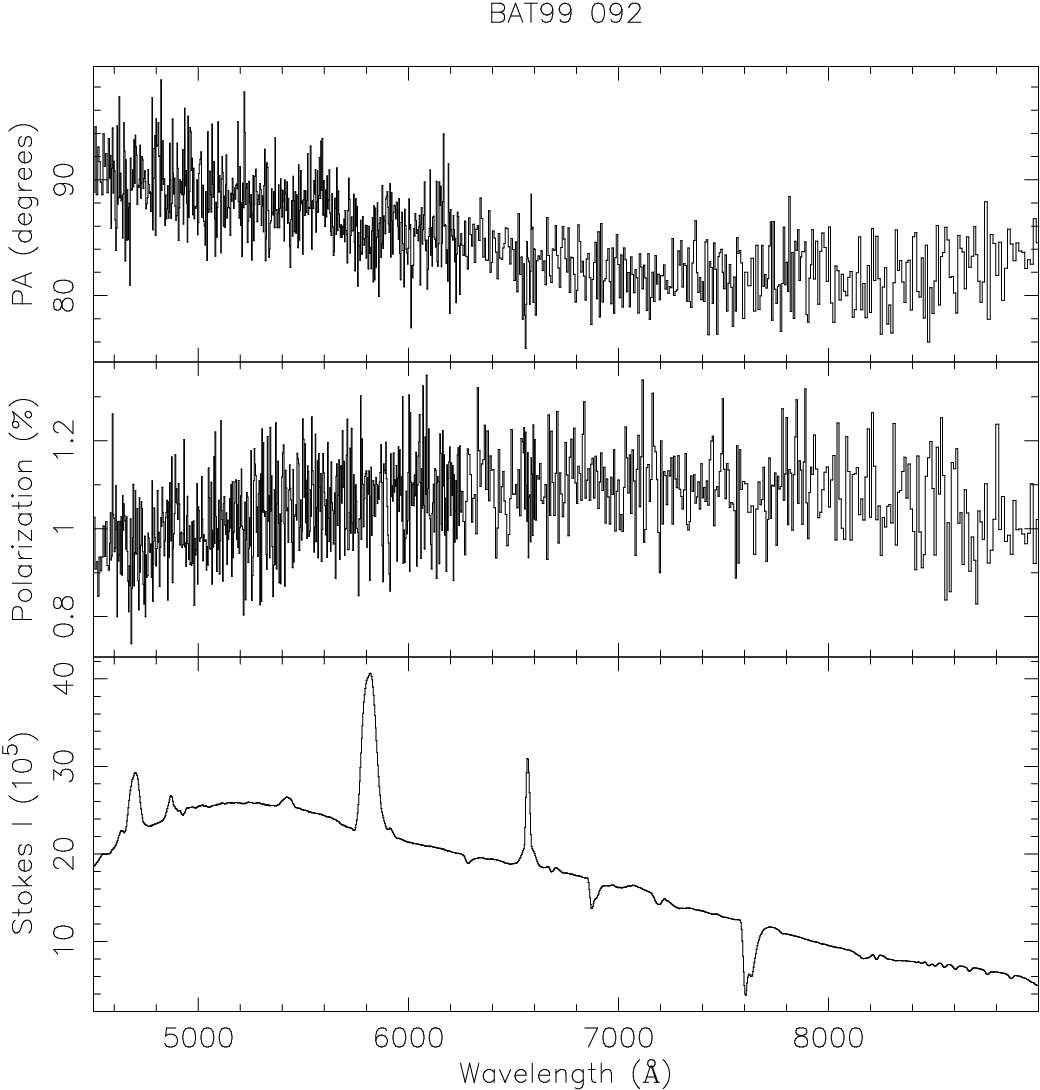}
\epsfxsize=0.25\textwidth\epsfbox{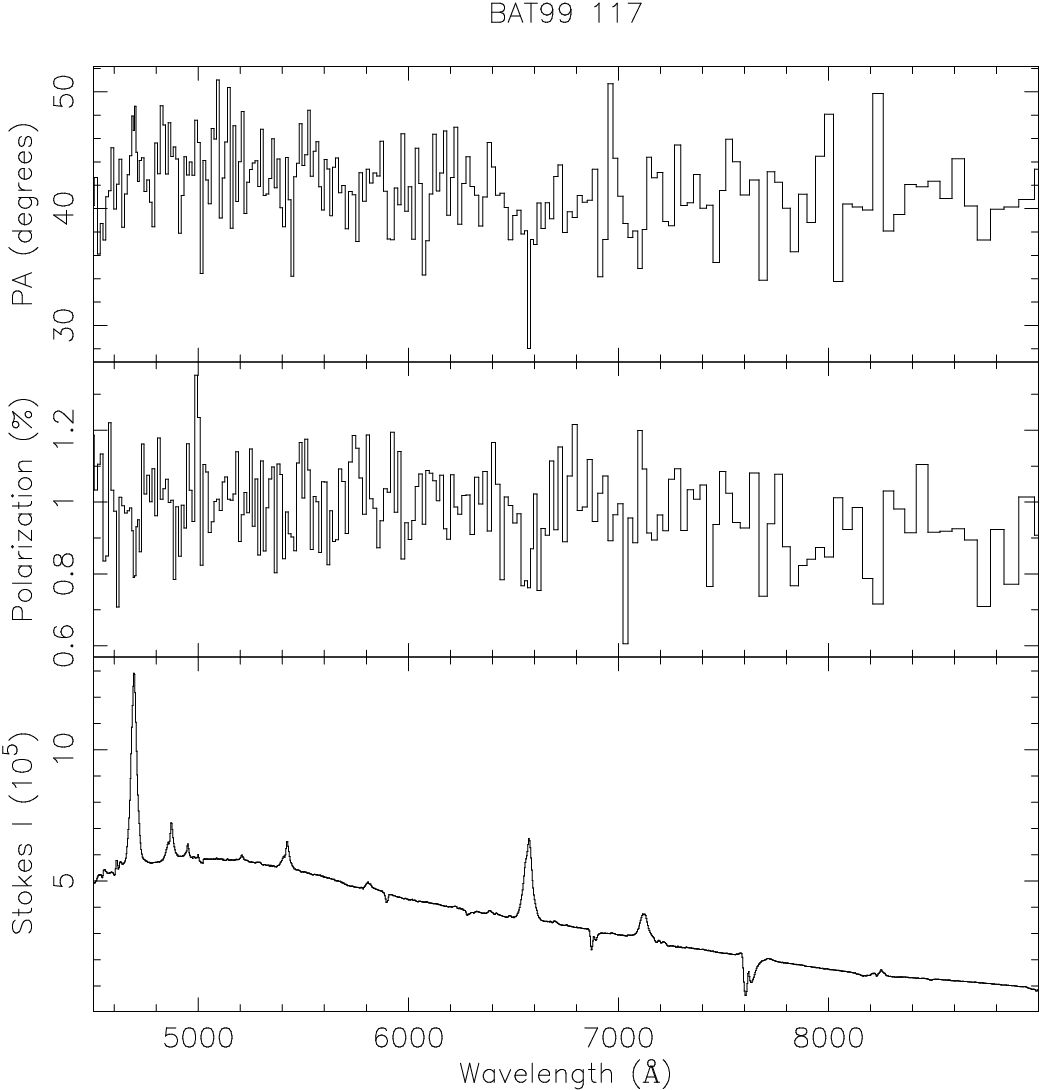}
\epsfxsize=0.25\textwidth\epsfbox{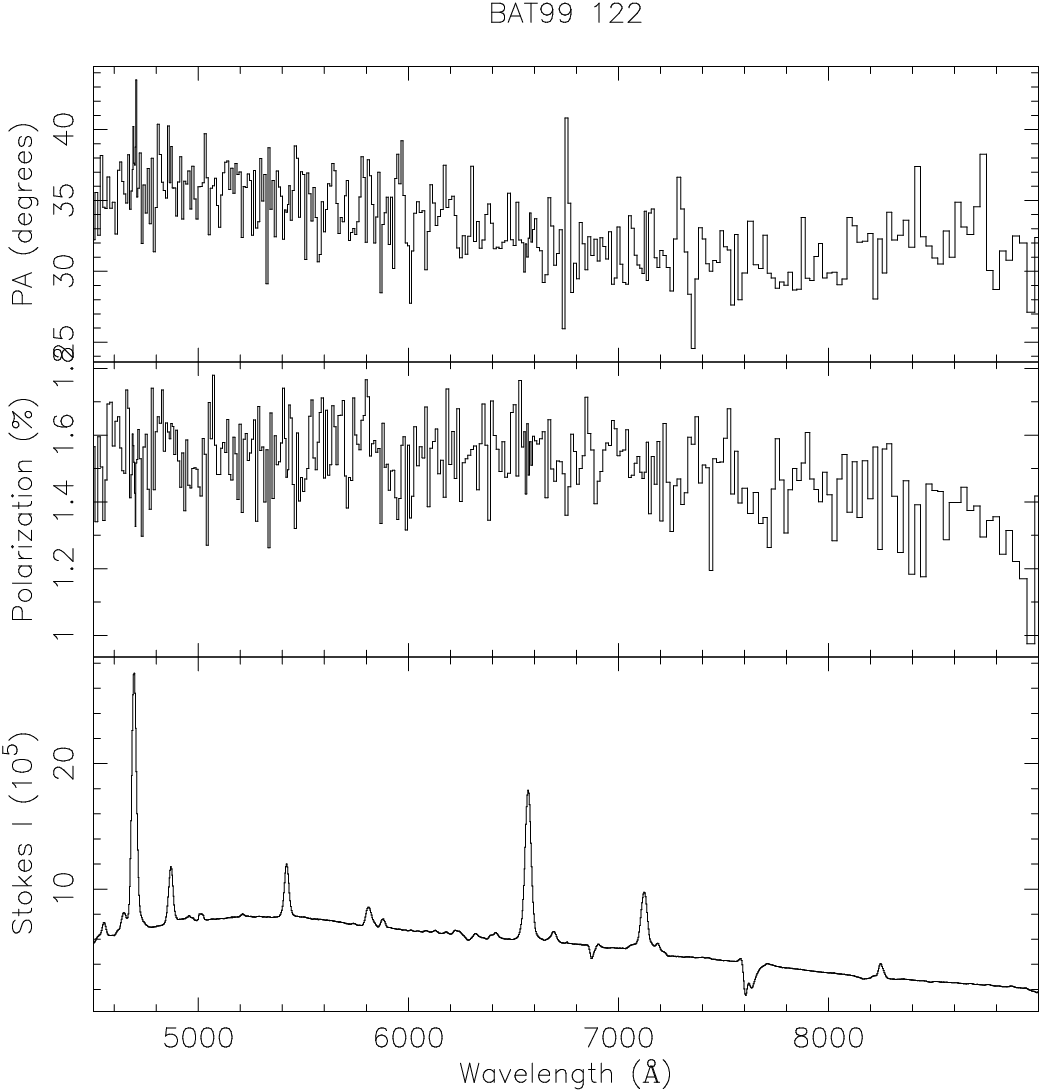}
\epsfxsize=0.25\textwidth\epsfbox{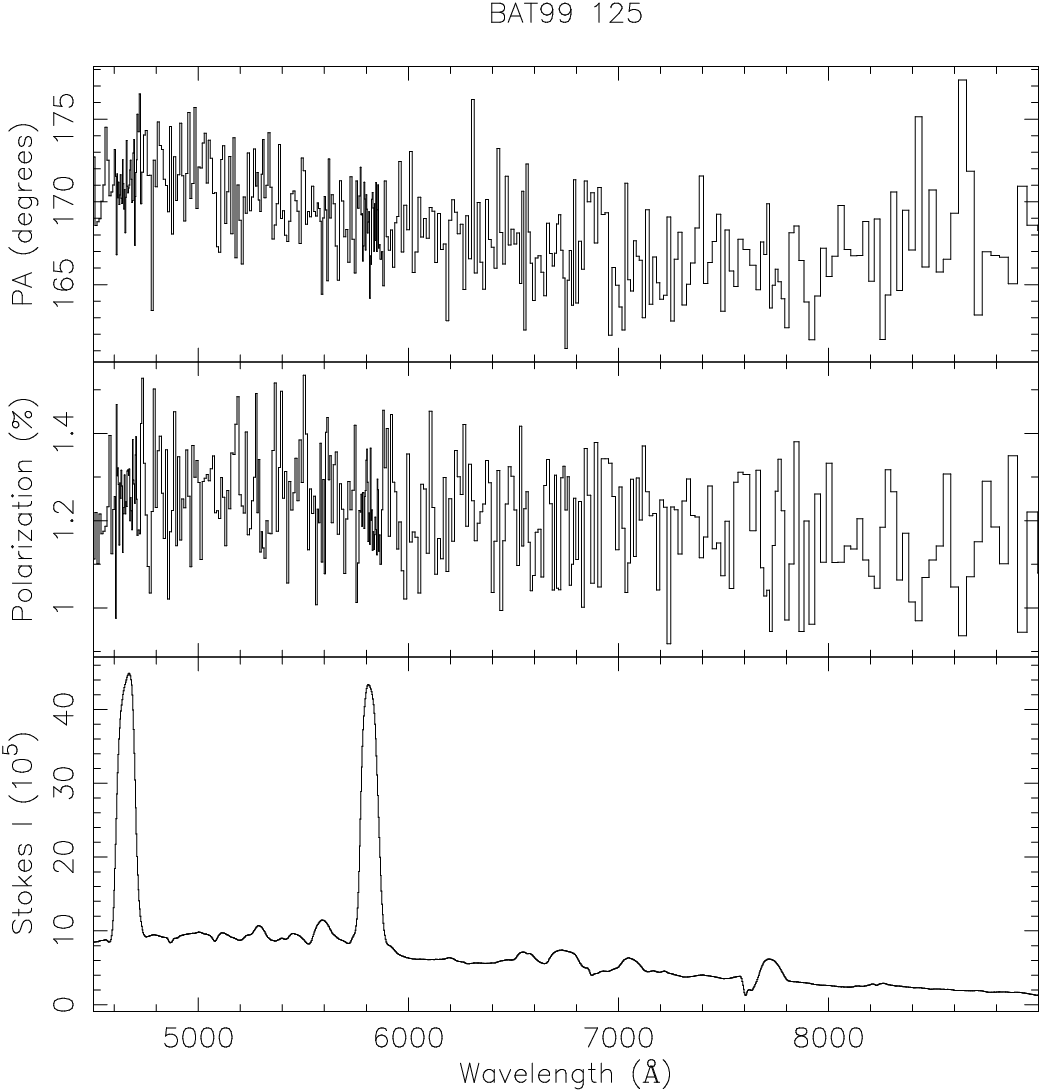}
}
\mbox{
\epsfxsize=0.25\textwidth\epsfbox{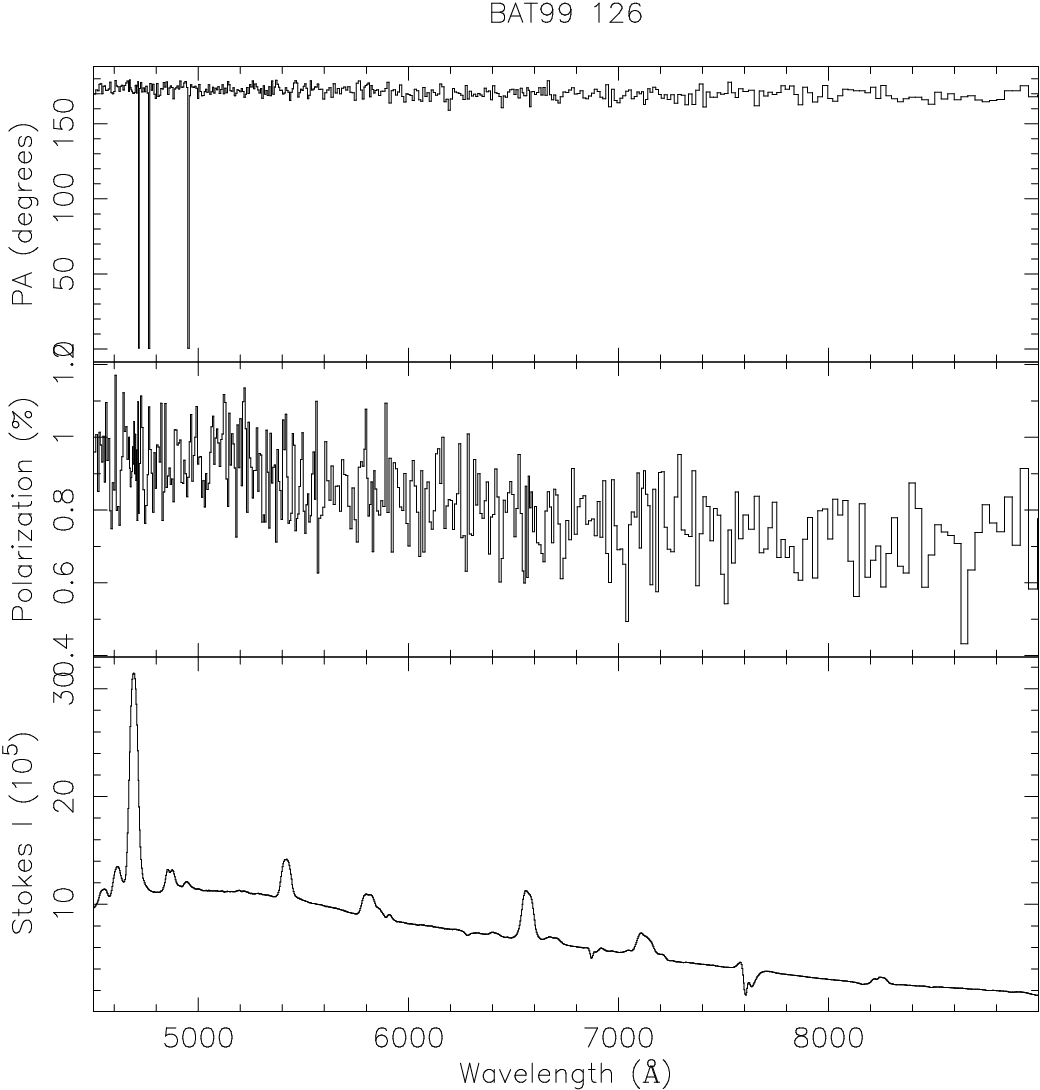}
\epsfxsize=0.25\textwidth\epsfbox{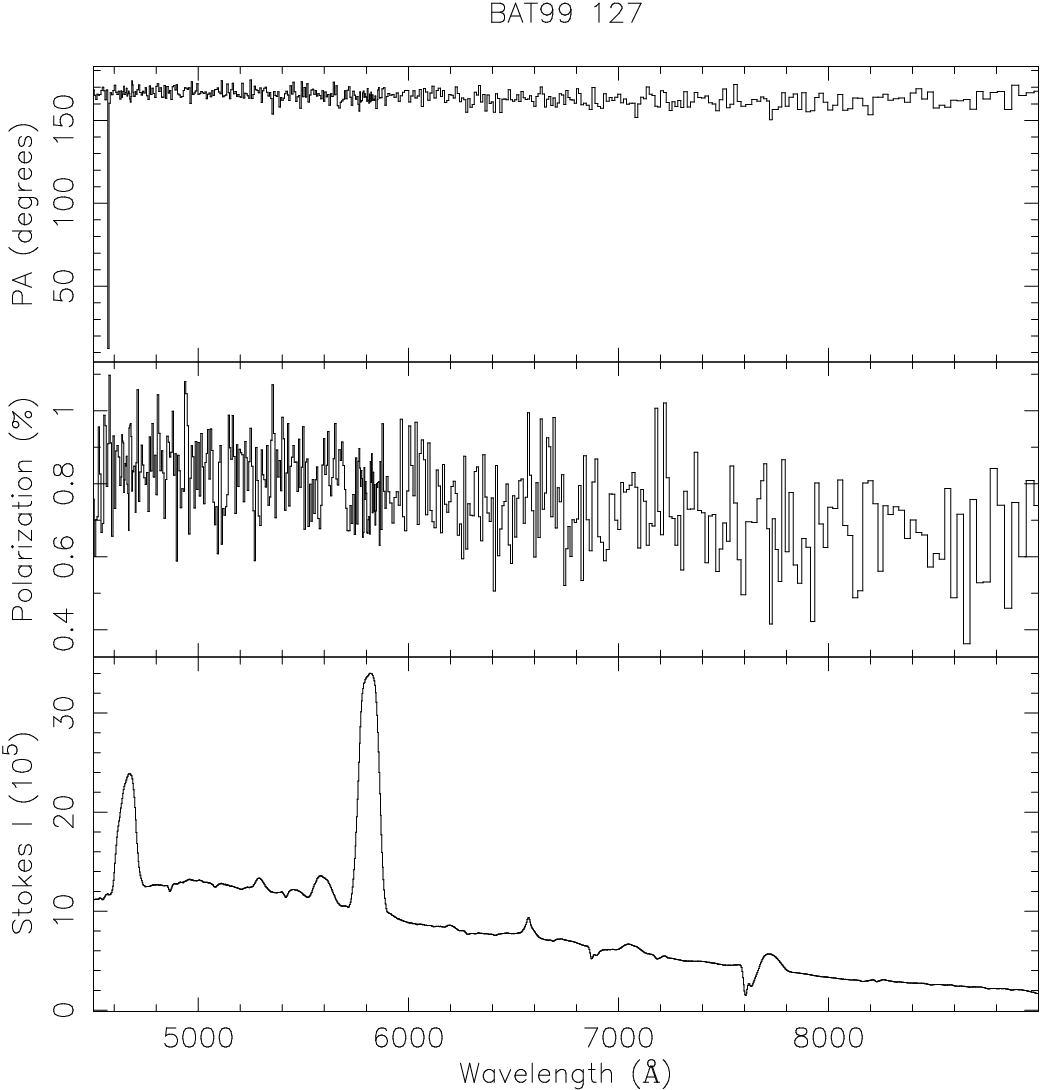}
}
\caption{Polarization spectra of the LMC WR objects {\it continued}.
\label{f_lmc2}}
\end{figure*}

The data for the 26 LMC WR stars are plotted in Figs.\,\ref{f_lmc} and\,\ref{f_lmc2}, and 
the continuum measurements are listed in Table\,1. The earlier measurements 
on the brighter LMC stars from Vink (2007) are added at the bottom of the Table.
Regarding continuum measurements, a significant body of previous data is available in
Matthewson \& Ford (1970). For BAT99-61, BAT99-84, BAT99-117, BAT-122, BAT99-28, BAT99-38, BAT99-119
our PA and the previous PA agree within the error bars. For the remainder of the sample the agreement is still present 
(within 2 or 3 $\sigma$), except for R99-59 and BAT99-22, where the PA is seen to vary significantly. 
Perhaps these two objects are intrinsically polarized. However, whilst BAT99-22 does show a 
line effect (see below), BAT99-59 does not.
Note that significant variations in the continuum PA might be interpreted as being the result of 
wind clumping (Davies et al. 2005).

We next consider the polarization and PA in lines with respect to continuum levels.   
New line effects are detected in BAT99-43 and BAT99-70.
Interestingly, as BAT99-43 is a WN star, whilst BAT99-70 is a WC star. So, it appears 
that line effects here are not confined to the WN spectral class. This 
situation is similar to that of the Milky Way, where a line effect was detected 
in the WC binary WR 137 (e.g. Harries et al. 1998), although the majority of detections 
was found amongst WN stars (Harries et al. 1998), coinciding with the 
presence of ejecta nebula (Vink et al. 2011).

\subsection{Binarity}

One of the aims of our study is to find out whether binarity is a necessary condition
for the detection of line-effects in WR stars, as for instance the 
famous Galactic line-effect WC star WR 137 is a dust-producing binary. 
There are basically two distinct manners in which binarity might 
affect the detection of line effects.

The first way involves intra-binary scattering in close binaries (see e.g. the recent
work by Hoffman \& Lomax 2015 on V444\,Cygni). The second way is more indirect, with 
binary evolution possibly leading to more rapid rotation 
of the WR star (e.g. Cantiello et al. 2007), which may then result in a line-effect
in an identical way as would be the case for a rotating single WR star. 
Spectropolarimetric monitoring might be a useful tool to distinguish between these options.

We first wish to find out if binarity is a prerequisite for line-effects 
amongst WR stars. For this reason, we correlate the incidence of line effects 
(column 9 in Table\,1) with binarity (column 10). To first order, there is no notable 
correlation between the two, which may indicate that binarity is not required 
to create line effects. It is of course possible that single stars have had a binary 
past (e.g. de Mink et al. 2014; Justham et al. 2014), so we cannot exclude the possibility that 
binarity has played a role in the physics that causes the enhanced rotation. 
Nonetheless, the lack of correlation between line effects and binarity suggests 
that line-effects in WR stars are not due to intra-binary scattering, thereby leaving stellar rotation as the most 
likely culprit. 

It has been noted (St-Louis 2013) that many of the line-effect WR stars also 
exhibit co-rotating interaction regions (CIRs), and that perhaps it is these spiral 
variable structures that cause (or contribute to) the polarization rather than an axi-symmetric
flattened wind. Harries (2000) performed Monte Carlo radiative transfer 
computations of spiral structures and concluded this to be unlikely. Nevertheless, whether the 
line effects in WR stars are caused by flattened winds or CIRs (Ignace et al. 2009) in both cases one would 
link the line-effects to stellar spin. I.e. line-effect stars are Wolf-Rayet stars 
with enhanced stellar rotation (Harries et al. 1998).

\begin{table*} 
\caption{Stellar parameters for WN stars with a line effect.
Parameters for the LMC WN stars are from Hainich et al. (2014), whilst the parameters for 
HD\,5980 are from Shenar et al. (2016). The Galactic WN parameters are from Hamann et al. (2006). 
Masses $M_\star$ and Eddington factors $\Gamma_{\rm e}$ are obtained from the mass-luminosity relations for both 
homogeneously H-burning and He-burning by Gr\"afener et al. (2011). 
The spectroscopically determined mass-loss rates $\dot{M}$ are scaled to a wind clumping factor $D=10$.
See the discussion in Vink et al. (2011) on the inclusion of WR\,136 as a line-effect object.
\label{tab:HRD}}
  \begin{tabular}{llllllllrr} \hline \hline
    \rule{0cm}{2.2ex}WR & WR sub-type & $T_\star$ & $X_{\rm H}$ &  $\log L$  &  $R_\star$  &  $M_\star$  & $\Gamma_{\rm e}$ & $\log(\dot{M})$ & $v$\\
    & [kK]      & &             & [$L_\odot$] & [$R_\odot$] & [$M_\odot$] & & [$M_\odot/{\rm yr}$] & [km/s] \\
\hline
\\
{\it SMC}:\\
\\
HD\,5980       & WN5 & 45 & 0.25 & 6.35    &  24    & 88.4 & 0.48       & -4.5 & 2200\\
               &     &    &      &         &        & 54.5 & 0.79       &      & \\
\hline
\\
{\it LMC}:\\
\\
BAT99-22       & WN9h & 32 & 0.4  & 5.75    &  25.1  & 43.4  & 0.28       & -4.85 & 400\\
               & &    &      &         &        & 21.5  & 0.56       &       & \\  
BAT99-33       & Ofpe/WN9 & 28 & 0.2  & 6.50    &  74.8  & 114.3 & 0.51       & -4.43 & 400\\
               &      &    &      &         &        &  72.2 & 0.81       &       & \\     
BAT99-43       & WN4o$+$OB & 67 & 0.0  & 5.85    &  6.3   & 27.0  & 0.40       & -5.15 &1600\\
               &      &    &      &         &        & 24.8  & 0.44       &       & \\
\hline \rule{0cm}{2.2ex}%
\\
{\it Milky Way}:\\
\\
    6         & WN4 & 89.1 & 0.0 & 5.6 & 2.65 & 17.9 & 0.342 & -4.5 & 1700\\
    16        & WN8 & 44.7  & 0.25 & 6.15   &  19.9  & 40.1 & 0.678 & -4.5 &650\\
              &     & 41.7  & 0.23 & 5.68   &  12.3  & 20.0 & 0.453 & -4.8 & \\
    40        & WN8 & 44.7  & 0.23 & 6.05   &  17.7  & 34.3 & 0.618 & -4.3 & 650\\
              &     & 45.0  & 0.15 & 5.61   &  10.6  & 18.2 & 0.397 & -4.5 & \\
    134       & WN6 & 63.1  & 0.0  & 5.6    &  5.29  & 17.9 & 0.342 & -4.6 & 1700\\
    136\footnote{} & WN6 & 70.8  & 0.12 & 5.4    &  3.34  & 13.7 & 0.316 & -4.7 & 1600\\
    \hline
  \end{tabular}
\end{table*} 

\section{Discussion}
\label{s_disc}

The incidence of line effects amongst LMC and SMC WR stars is low, which is 
unlikely due to their apparent faintness. Harries et al. (1998) noted that Galactic 
WR stars that show a line effect are polarized at levels $>$ 0.3\%, arguing
that polarized WRs are relatively rapid rotators. 
The S/N ratios achieved in the present study of MC WRs 
are large enough to detect similarly-sized line effects.

Amongst the LMC WR sample, 2 out of 26, have line effects in addition to 
the 2 out of 13 from the brighter LMC Vink (2007) WR sample. 
This brings the overall incidence amongst the 
LMC sample to 4 out of 39, i.e. not significantly 
different from the 15-20\% in the similarly-sized Galactic sample (of 29) 
of Harries et al. (1998). Our SMC results of 1 in 12 comply as well. 
On the basis of the low incidence of line-effects amongst the 
bright LMC sample Vink (2007) speculated his results could be explained 
if rapidly rotating WRs as progenitors for GRBs were confined to metallicities 
below that of the LMC (see also Wolf \& Podsiadlowski 2007 on entirely different grounds). 
We can no longer draw this conclusion, as the incidence of line-effect WR stars in the 
SMC, with a $Z$ notably smaller than that of the LMC and Galaxy, is similarly low. 

When we split the WR sample into WC stars separately, we note that the frequency 
of Galactic WC stars of 1/8 does not increase either. In fact, for the LMC sample 
the frequency is 1/16. The lack of an increase in WR polarization
at lower $Z$ challenges the rotation-induced CHE model for GRB production.

\subsection{The challenge to rotation-induced CHE for making GRBs}
\label{sec_cha}

The most popular model for the production of GRBs is that of rotation-induced  
CHE (Yoon \& Langer 2005; Woosley \& Heger 2006; Levan et al. 2016). The attraction
of the model is that rapid rotation mixes the star completely, which avoids the 
expansion of the star towards the red side of the HRD. The blue-ward 
evolution for the magnetic models (Brott et al. 2011) in combination with smaller 
WR mass loss at lower metal content (Vink \& de Koter 2005)
could ensure the maintenance of rapid rotation until collapse.

Evidence for CHE in the low-$Z$ environment of the SMC was presented by
Martins et al. (2009) for SMC 1 and 2. 
The evidence for CHE was interpreted as being due to rapid stellar rotation.  
In the present study, line effects were not detected for these SMC objects, and it would be 
unlikely that both of them are observed pole-on. 
It is therefore more likely that line-effect WR stars are special for reasons other 
than rotationally-induced CHE. Vink et al. (2011) discovered a highly significant correlation 
between Galactic line-effect WR stars 
and WR stars with ejecta nebulae (see below). 

Furthermore, another issue has come to light. In the original rotation-induced 
CHE models of Yoon \& Langer (2005) GRB production was predicted to take place 
for initial masses as low as 20$M_{\odot}$, but observations of GRB host galaxies 
prefer significantly higher initial masses, above 40$M_{\odot}$ (Raskin et al. 2008; Levan et al. 2016),
as GRBs are more concentrated on the brightest regions of their host galaxies than supernovae, 
hinting at more massive stars (Fruchter et al. 2006).

\subsection{Alternative: post-LBVs} 

For the Galactic sample, the high coincidence between line-effect WR stars and those 
WN stars with ejecta nebula suggested that the rotating WR subset are post-LBVs. 
Moreover, for the case of weak coupling between the stellar core and 
envelope, i.e. without the presence of a magnetic field 
(Hirschi et al. 2005; Petrovic et al. 2005; Georgy et al. 2013; Groh et al. 2013), the WR rotation
rates inferred (see e.g. Chen\'e \& St-Louis 2010) are sufficiently large to fulfill the 
angular momentum conditions for GRB production (Vink et al. 2011; Gr\"afener et al. 2012).
Whilst these Galactic line-effect WN stars (see bottom of Table 2) have the 
correct ingredients, they may not be GRB progenitors due to their high metallicity.

In the current low-$Z$ study, we found 3 line-effects in LMC WN stars.
As was already noted, from the current study BAT99-70 is a WC star, whereas BAT99-43 is of WN type. 
The two line-effect stars from Vink (2007) are both very late-type WNL, WN9 objects. 
When considering the parameters of a very large sample of over 100 
LMC WN stars of Hainich et al. (2014), 
the line-effect stars BAT99-22 and BAT99-33 do indeed stand out, as these late-type WN objects 
have far slower outflow velocities (only of order 400 km/s) than any of the other 
WR stars (with terminal wind velocities of thousands of km/s). 

A potential reason for this could be that these winds are aspherical and 
observed close-to edge-on. Asphericity could be due to wind-compression 
(Bjorkman \& Cassinelli 1993; Ignace et al. 1996) or non-radial line forces
(Owocki et al. 1996). However, 2D wind hydrodynamics calculations 
show that even for very rapid rotation one would only expect relatively small
changes in the mass-loss rate, terminal wind velocity, and wind density
as a function of latitude (M\"uller \& Vink 2014). It is therefore more likely
that the slow wind features in  BAT99-22 and BAT99-33 are intrinsic to
the type of star.

Interestingly, both BAT99-22 and BAT99-33 have in the past been classified as Ofpe/WN9 stars (Bohannan \& Walborn 
1989), and these objects are thought to be closely related to LBVs, esp. since the Ofpe/WN9 star
R\,127 went into outburst (Stahl et al. 1983; Pasquali et al. 1997; Vink 2012). 
Furthermore, Nota et al. (1995) found that five of 
the Ofpe/WN9 stars show the presence of nebular emission lines, indicating a surrounding 
nebulosity. 

There thus seems to be a coherent picture developing with respect to the LMC Ofpe/WN9 stars: they 
have slow outflow velocities, CS nebulae, and are predominately linearly polarized. 
One reason for the larger incidence of line effects amongst these later type WN stars
 could be that it is easier to detect asymmetries in a slower wind, 
as polarization levels seem to increase for larger stellar radii 
(Robert et al. 1989; Davies et al. 2005). However, this would not explain 
the correlation between line-effect WR stars and 
stars with ejecta nebula found by Vink et al. (2011).
That correlation was interpreted as a link between rapid rotation and age. 
Young WR stars that have only recently transitioned from the RSG or LBV phase
may still maintain their rotation rates. From photometric periods 
Gr\"afener et al. (2012) estimated moderate rotation speeds of 36...120 km/s for line-effect
WR stars. Assuming a fixed $\Omega = v_{\rm rot} / R_\star$ it is clear that the evolutionary 
progenitor radius cannot be much larger. 
For S\,Dor LBVs, such as AG\,Car and HR\,Car, $v_{\rm rot}$ is of order 
200 km/s (Groh et al. 2009). With LBV radii of order 70\,$R_{\odot}$, the associated late-type 
Galactic WN8 stars WR\,16 and WR\,40 with radii up to 20\,$R_{\odot}$ are rather consistent
with an LBV - WR evolutionary transition, whilst the large radii of Magellanic Cloud WR stars
make the post-LBV scenario even more likely. Assuming a fixed $\Omega$, a 
post-RSG scenario -- with extremely large radii of $>>$ 100\,$R_{\odot}$ -- 
would however require extremely rapidly rotating RSGs, which are not known.

Smith et al. (2004) noted that post-RSG/LBV stars when 
they return from the red part of the HRD (where they may bounce several times against the cool side of the 
bi-stability jump\footnote{The bi-stability jump is a strong mass-loss discontinuity at 
$\sim$20\,000\,K (Petrov et al. 2016)}) 
to the blue part of the HRD (on the hot side of the bi-stability jump) might appear as Ofpe/WN9 stars 
with slowly expanding ring nebulae. If the Ofpe/WN9 stars and line-effect objects 
may indeed be identified as post-RSG or post-LBV objects with a nebula, it is likely they 
would have lost a significant fraction of their mass, which will 
bring them closer to their Eddington limit (Humphreys \& Davidson 1994; Vink 2012). 
An object in close proximity to the Eddington limit would not only be subjected 
to a larger mass-loss rate (Castor et al. 1975; Vink \& Gr\"afener 2012), but also a slower
wind velocity (Castor et al. 1975; Gr\"afener \& Hamann 2008), which might 
be the reason why Ofpe/WN9 stars like BAT99-22 and BAT99-33 have such slow 
terminal velocities, of just 400 km/s. 

Can we link these WNL stars to the Galactic line-effect WN stars? 
Looking at the bottom of Table 2 it might be interesting to note that two of the five 
are the runaway WN8 stars, WR 16 and WR 40 (Walborn \& Fitzpatrick 2000), with slow 
outflow speeds of order 650 km/s. 
For decades it has been known that Galactic WN8 stars are 
special (Moffat \& Shara 1986; Crowther et al. 1995), in terms of their larger photometric variability, 
their association with ejecta nebulae, their low binary frequency, and relatively ``isolated'' 
spatial distribution. In this sense they share many of these properties 
with LBVs (see Smith \& Tombleson 2015). Perhaps both their isolation (Kenyon \& Gallagher 1985) 
and their low-binary frequency (Vink 2012) indicate a binary 
past.

In any case, taking the special properties of both LMC and Milky Way WNL stars together, the line-effect 
results indeed suggest a link between stellar rotation and the presence of a nebula. 
For LBVs,
Gvaramadze \& Kniazev (2016) suggested that isolated/runaway LBV stars are more likely
surrounded by a CS nebula than for cluster-stars due to the potential destructive effects of 
neighboring massive stars, but this would not explain {\it why} isolated stars would rotate faster 
than clustered objects. 
It is therefore more likely there is a {\it physical} connection between 
stellar rotation and the presence of a nebula.

Next, we will search for a possible link between 
stellar rotation of GRB progenitors and the connection to wind/nebula features in 
afterglow spectra.

\subsection{CS features in GRB afterglow spectra}

Absorption features in GRB afterglow spectra have been 
interpreted as either due to the absorption in the intervening 
interstellar medium (ISM), or the direct 
circumstellar medium (CSM). Interest was 
triggered by the case of GRB 021004 (M\"oller et al. 2002; Schaefer et al. 2003; Mirabal et al. 2003;
Fiore et al. 2005; Starling et al. 2005; Castro-Tirado et al. 2010), which showed absorption features
both at intermediate ($\sim$ 500 km/s) and high velocity 
($\sim$ 3000-4000 km/s). Whilst the high velocity features could either be explained by the 
wind outflow speed of a classical WR star, or acceleration by the burst's radiation, the 
intermediate velocity of order $\sim$ 500 km/s has been more challenging to explain. 

Van Marle et al. (2005) performed hydrodynamical simulation of the evolution of the 
circumstellar medium around post-RSG/LBV stars that could potentially explain intermediate 
velocities, but WR nebula expansion speeds are far lower (Marston 1995) than predicted 
by van Marle et al (2005). 

If our post-LBV scenario for GRB progenitors that so nicely explains the correlation 
between stellar rotation and the presence of ejecta nebulae is correct, it might be relevant
that the outflow speeds in both the LMC WN9/Ofpe and the Galactic WN8 stars are 
of order 500 km/s, possibly accounting for intermediate velocities seen in GRB afterglow
spectra. 

Finally, we note that although Chen et al. (2007) did not find any 
evidence for CS absorption features in the range $-$1000 to $-$5000 km/s within 
a sample of 5 GRBs, Fox et al. (2008) preferred a CS origin for at least 
some of their sample of 7.

\subsection{The SMC WR/LBV HD\,5980: CHE after all?}

Whilst the late-type WN LMC stars BAT-22 and BAT-33 might be explained by the post-LBV 
scenario, BAT-43 is earlier with a much smaller radius and a 
higher effective temperature (see Table 2). For this object some form of CHE might still be a more suitable scenario. 
CHE might no longer be confined to low $Z$, as for instance Martins et al. (2013) 
suggested cases at Galactic and LMC metallicity. Even for the very massive star (VMS) 
WN5h star VFTS\,682, Bestenlehner et al. (2011) suggested that the high stellar 
effective temperature would be best explained by CHE. 
One should note that rapid rotation is not an absolute requirement for CHE, but it 
can also be caused by a large convective core and high mass-loss rates 
for VMS (Gr\"afener et al. 2011; Yusof et al. 2013; Vink 2015; K\"ohler et al. 2015).

The SMC binary system HD\,5980 was suggested to be the result of CHE, making it a 
potential GRB progenitor (Koeningsberger et al. 2014). Our detection of a line-effect
in HD\,5980 might thus be relevant. 
Given the challenges for rotation-induced CHE (Sect.~\ref{sec_cha}), it seems 
more likely that CHE in HD\,5980 is not due to rotation-induced CHE, but the result of 
VMS-specific physics of a large convective core and strong mass loss instead. 
Note that all line-effect WN stars in Table 2 have luminosities 
$\logl$ $\ge$ $5.4$ (see Table\,2), corresponding to initial masses above
the required 40$M_{\odot}$ for GRB production (Levan et al. 2016). 
 
\section{Summary}
\label{s_sum}

The detection of gravitational waves from a merging ``heavy'' black hole associated 
with GW\,150914 has created a renewed interest in 
the formation of stellar- mass black holes. One of the largest surprises was the 
very large mass of these objects, hinting at a low $Z$ environment. In turn, 
lower $Z$ may lead to less spin-down, and subsequently larger WR spin rates. 
The rotation rates of WR stars and black hole progenitors have thus become 
key aspects for our understanding of massive single and binary star evolution 
towards collapse.
The most popular scenarios involve classical evolution through mass loss (e.g. Belczynski et al. 2016) 
and rapid rotation with quasi-CHE (e.g. de Mink \& Mandel 2016; Marchant et al. 2016). 

To obtain empirical constraints on black hole progenitor spin 
we measured wind asymmetries in all 12 known WR stars in the SMC at $Z=1/5\zsun$, and also  
within a significantly enhanced sample of objects in the LMC at $Z=1/2\zsun$, tripling the previous  
sample of Vink (2007). 
This total LMC sample size of 39 made it appropriate for comparison to the Galactic sample.
We measured wind asymmetries with 
linear spectropolarimetry and we detected new line effects in the 
LMC WN star BAT99-43 and the WC star BAT99-70, as well as 
HD\,5980 in the SMC, which we confirm to be evolving chemically homogeneously. 

With previous reported line effects in the late-type WNL (Ofpe/WN9) 
objects BAT99-22 and BAT99-33, this brings the total LMC 
WR sample to 4, i.e. a frequency of $\sim$10\%, which was 
not found to be higher than amongst the Galactic WR sample.
As WR mass loss is likely $Z$-dependent, our Magellanic Cloud line-effect WR stars 
may maintain their surface rotation and fulfill the basic conditions for producing 
long GRBs.

The similar fraction of line-effect WR stars at low $Z$ in comparison to the Milky Way 
casts doubt on the rotationally-induced CHE scenario for producing GRBs and 
objects like GW\,150914 (see Sect.~\ref{sec_cha}). Instead, our data seem consistent with a 
post-LBV channel, as well as resulting from CHE due to physics specific to VMS, i.e.
a large convective core and strong mass loss.

\begin{acknowledgements}
We would like to thank the anonymous referee for their constructive comments. 
This research made use of the SIMBAD database, operated at CDS, Strasbourg, France.
JSV is funded by the Northern Ireland Department of Communities (DfC). TJH was funded 
by ST/M0012174/1.

\end{acknowledgements}

\end{document}